\documentclass[preprint,authoryear,12pt]{elsarticle}

\usepackage{latexsym}
\usepackage{amssymb}
\usepackage{mathrsfs}
\usepackage{amsmath}
\usepackage{epsfig}
\usepackage{graphicx}
\usepackage{enumerate}
\usepackage{bm} 
\usepackage{bbm} 
\usepackage{booktabs}
\usepackage[hang]{subfigure} 

\usepackage{rotating}
\usepackage{lscape}
\usepackage[latin1]{inputenc}

\usepackage[
setpagesize=false, 
pdfborder={0 0 0}, 
pdfpagemode=UseOutlines, 
]{hyperref} 

\newcommand{\bbr}{\mathbb{R}}
\newcommand{\bbc}{\mathbb{C}}
\newcommand{\bbh}{\mathbb{H}}

\newcommand{\bx}{\boldsymbol{x}}
\newcommand{\btheta}{\boldsymbol{\theta}}
\newcommand{\bX}{\boldsymbol{X}}
\newcommand{\bu}{\boldsymbol{u}}
\newcommand{\bU}{\boldsymbol{U}}

\newcommand{\calT}{{\cal T}}
\newcommand{\calt}{{t}}
\newcommand{\cali}{{\cal I}}

\newcommand{\calk}{{\cal K}}

\usepackage[
thmmarks, 
amsmath, 
hyperref 
]{ntheorem} 

\theoremstyle{break} 
\theoremheaderfont{\bfseries}
\theorembodyfont{\itshape}
\theoremseparator{}

\newtheorem{proposition}{Proposition}
\newtheorem{corollary}{Corollary}

\newtheorem{example}{Example}
\theoremstyle{nonumberbreak} 
\theoremsymbol{$\Box$}

\journal{Insurance: mathematics \& economics}

\begin{document}

\begin{frontmatter}

\title{A goodness-of-fit test for regular vine copula models}
\author{Ulf Schepsmeier }
\address{Center for Mathematical Sciences, \\
Technische Universit\"at M\"unchen,\\
Boltzmannstraße 3, 85748 Garching b. München, Germany,\\ 
email: \texttt{schepsmeier@ma.tum.de},
phone: +49 89 289 17422.}

%
%
%

\begin{abstract}
We introduce a new goodness-of-fit test for regular vine (R-vine) copula models. R-vine copulas are a very flexible class of multivariate copulas based on a pair-copula construction (PCC). The test arises from the information matrix equality and specification test proposed by \cite{White1982} and extends the goodness-of-fit test for copulas introduced by \cite{HuangProkhorov2011}. 
The corresponding critical value can be approximated by asymptotic theory or simulation. The simulation based test shows excellent performance with regard to observed size and power in an extensive simulation study, while the asymptotic theory based test is inaccurate for $n\leq 10000$ for a 5-dimensional model (in $d=8$ even 20000 are not enough). The simulation based test is applied to select among different R-vine specifications to model the dependency among exchange rates.
\end{abstract}


\end{frontmatter}

\section{Introduction}
\label{sec:introduction}

\noindent
Goodness-of-fit tests are used to verify a statistical model. 
In the literature one can find a number of copula goodness-of-fit tests \citep{Genest_Remillard_2,Genest2009,Berg,HuangProkhorov2011,Dufour2012}, but most of them are designed for bivariate copula families or only implemented and tested for the bivariate case. 
Some of them claim to work for higher dimensional copulas as well. Further, only a few are "blanket", i.e.~having desirable properties such as no requirement for "parameter tuning or other strategic choices" \citep{Genest2009}. Most of them are based on the empirical copula or on Kendall's or Rosenblatt's probability integral transformation (PIT).\\
In this paper we propose a rank-based "blanket" goodness-of-fit test for a special class of high dimensional copulas - the vine copulas, based on the information matrix equality and specification test proposed by \cite{White1982}. It extends the goodness-of-fit test of \cite{HuangProkhorov2011} designed for (bivariate) copulas to the vine copula class.\\
Because of its high flexibility and its easy construction R-vines became very popular in the recent years and many authors added inference, selection methods and algorithms \citep{Aas_Czado,Czado,MinCzado2010,min:czado:2010:scomdy,BrechmannCzadoAas2012,CzadoSchepsmeierMin2011}, asymptotic theory \citep{haff2010,StoeberSchepsmeier2012,DissmannBrechmannCzadoKurowicka2011}, software (CDVine by \citealt{BrechmannSchepsmeier2011}, VineCopula by \citealt{VineCopula}), and applications \citep{BrechmannEuroStoxx2012}. But to our knowledge no goodness-of-fit test exists.\\ 
In \cite{Aas_Czado} a goodness-of-fit test based on PIT and a transformation introduced by \cite{Breymann2003} is suggested for vine copulas but not further studied or tested. They further state that an Anderson-Darling goodness-of-fit test may be applied for the subsequent independence test. On the other hand e.g.~\cite{Genest_Remillard_2} and references within mention that this approach has less power and does not even maintain its nominal level.
\cite{BergAas2009} apply two further approaches originally proposed for copulas to a 4-dimensional vine copula. Their goodness-of-fit tests are based on the empirical copula $C_n$ and on Kendall's process $\calk_n = \sqrt{n}\{K_n-K_{\hat{\theta}_n}\}$, respectively, where
\[
	K_n(t) = \frac{1}{n+1}\sum_{j=1}^n 1_{\{C_n(\boldsymbol{U}_j)\leq t\}}\qquad\text{and}\qquad K_{\hat{\theta}_n}(t) = P(C_{\hat{\theta}_n}\leq t)
\]
are the empirical distribution function of $C_n(u)$ and $K_n(t)$ the parametric estimate of Kendall's dependence function $K(t)$, respectively. Further, $\boldsymbol{U}_j = (U_{j1},\ldots,U_{jd})^T$ are the $U(0, 1)^d$ copula data and $\hat{\theta}_n$ is the maximum likelihood estimator corresponding to the parametric copula $C_{\theta}$. Both tests use the Cramér-von Mises statistic with $C_n$ and $K_n$, respectively. Again, the tests are not studied in detail for the vine copula case. Furthermore, in general $C_{\hat{\theta}_n}$ and $K_{\hat{\theta}_n}$ can not be determined analytically for vine copulas but only via simulation.\\  
Model verification methods for vine copulas are usually based on the likelihood. AIC or BIC are the classical comparison measures, in addition to the likelihood, which ignores model complexity. For vine copulas tests proposed by \cite{Vuong} and \cite{Clarke}, suitable for non-nested models, are for example used in \cite{CzadoSchepsmeierMin2011}. Both test statistics involve likelihood ratios of two specified (vine) models with estimated parameters. However, goodness-of-fit tests for vine copula models verifying the chosen pair-copula families have not yet been investigated. The main contribution is to fill this gap.\\
Especially a comparison with multivariate copulas such as the multivariate Gauss or Student's t copula can now be done in a more quantitative way based on p-values. \\
Further, the influence of model selection and estimation for the margins are usually not investigated in a copula goodness-of-fit test. Therefore no uncertainty in the margins is assumed. 
We show that given a plausible model for the margins our goodness-of-fit test still works reasonable given estimated margins.\\
The remainder of this paper is structured as follows: We introduce R-vines in more detail in Section \ref{sec:rvine}, while the general misspecification test for copulas based on the information matrix equivalence is explained in Section \ref{sec:test}. The specific R-vine goodness-of-fit test is then described in Section \ref{sec:gofRvine}. We investigate its size and power in Section \ref{sec:powerstudy} in an extensive simulation study. Section \ref{sec:extension} extends the proposed goodness-of-fit test to unknown margins. An application is given in Section \ref{sec:application} comparing different vine specifications for an 8-dimensional exchange rate data set. Section \ref{sec:discussion} gives conclusions and shows areas for further research.

\section{Introduction to R-vine copula models}
\label{sec:rvine}

\noindent
A special case of constructing a multivariate density with bivariate copulas was first discussed by \cite{Joe2}. \cite{Bedford_Cooke2001,Bedford_Cooke} independently constructed multivariate densities using $d(d-1)/2$ bivariate copulas, which are identified by nested trees $T_i=(V_i, E_i)$. Here $V_i$ denotes the nodes while $E_i$ represents the set of edges. This process was called by \cite{Aas_Czado} a pair-copula construction (PCC).
Following the notation of \cite{Czado} with a set of bivariate copula densities  $\mathcal{B} = \left\{c_{j(e),k(e)|D(e)} \vert e \in E_i, 1 \leq i \leq d-1 \right\}$ corresponding to edges $j(e),k(e)|D(e)$ in $E_i$, for $1\leq i\leq d-1$ a valid $d$-dimensional density can be constructed by setting
\begin{gather}
\begin{split}
	&f_{1,\ldots,d}(u_1,\ldots,u_d)\\
	& \ \ =\prod_{i=1}^{d-1}\prod_{e\in E_i}c_{j(e),k(e);D(e)}(F_{j(e)|D(e)}(u_{j(e)}|\boldsymbol{u}_{D(e)}),F_{k(e)|D(e)}(u_{k(e)}|\boldsymbol{u}_{D(e)})).
	\label{eq:density}
\end{split}
\end{gather}
Here $\boldsymbol{u}_{D(e)}$ denotes the subvector of $\boldsymbol{u}=(u_1,\ldots,u_d)\in[0,1]^d$ determined by the set of indices in $D(e)$, which is called the {\it conditioning set} while the indices $j(e)$ and $k(e)$ form the {\it conditioned set}. 
The required conditional cumulative distribution functions $F_{j(e)|D(e)}(u_{j(e)}|\boldsymbol{u}_{D(e)})$ and $F_{k(e)|D(e)}(u_{k(e)}|\boldsymbol{u}_{D(e)}))$ can be calculated as the first derivative of the corresponding copula cumulative distribution function (cdf) with respect to the second copula argument, i.e.
{\small
\begin{gather}
\begin{split}
	F(u_{j(e)}|\boldsymbol{u}_{D(e)}) &= \frac{\partial C_{j(e),j^{\prime}(e);D(e)\setminus j^{\prime}(e)}(F(u_{j(e)}|\boldsymbol{u}_{D(e) \setminus j^{\prime}(e)}),F(u_{j^{\prime}(e)}|\boldsymbol{u}_{D(e)\setminus j^{\prime}(e)}))}{\partial F(u_{j^{\prime}(e)}|\boldsymbol{u}_{D(e)\setminus j^{\prime}(e)})} \\
	&=: h_{j(e),j^{\prime}(e);D(e)\setminus j^{\prime}(e)}(F(u_{j(e)}|\boldsymbol{u}_{D(e)\setminus j^{\prime}(e)}),F(u_{j^{\prime}(e)}|\boldsymbol{u}_{D(e)\setminus j^{\prime}(e)})),
\end{split}
\label{eq:hfunction}
\end{gather}}\noindent
For regular vines there is an index $j^{\prime}(e)$ in the conditioning set of indices given by edge $e$, such that the copula $C_{j(e),j'(e);D(e)\setminus j'(e)}$ is in the set of pair-copulas $\mathcal{B}$ \cite{DissmannBrechmannCzadoKurowicka2011}. 
Doing so, we assume that the copula 
$ C_{j(e),j'(e);D(e)\setminus j'(e)}$ does not depend on the values $\boldsymbol{u}_{D(e)\setminus j^{\prime}(e)}$, i.e.~on the conditioning set without the chosen variable $u_{j^{\prime}}(e)$. This is called the {\it simplifying assumption}.
In the literature Expression (\ref{eq:hfunction}) is often denoted as a h-function.
A PCC is called an R-vine copula if all marginal densities are uniform. Given an i.i.d.~sample in $d$ dimensions of size $n$, denoted by $\bx_1,\ldots,\bx_n$, the unknown margins $F_i, i=1,\ldots,d$ can be estimated empirically. As proposed by \cite{Genest1995} these estimates can be used to transform the data $\bx$ to an approximate sample $\bu:=(\bu_1,\ldots,\bu_n)$ in the copula space. \\

\begin{example}[3-dim pair-copula construction]
Let $u_1=F_1(x_1), u_2=F_2(x_2)$ and $u_3=F_3(x_3)$, and call $u_1, u_2$ and $u_3$ copula data. Then the corresponding pair-copula construction under the simplifying assumption is
\begin{equation*}
	c_{123}(u_1,u_2,u_3) = c_{12}(u_1,u_2)  c_{23}(u_2,u_3)  c_{13;2}(C_{1|2}(u_1|u_2),C_{3|2}(u_3|u_2)). \\
\end{equation*}
where $C_{i,j}$ denotes the conditional distribution function of $U_i$ given $U_j$.\\
\end{example}

\begin{figure}[hbt]
	\centering
		\includegraphics[width=0.75\textwidth]{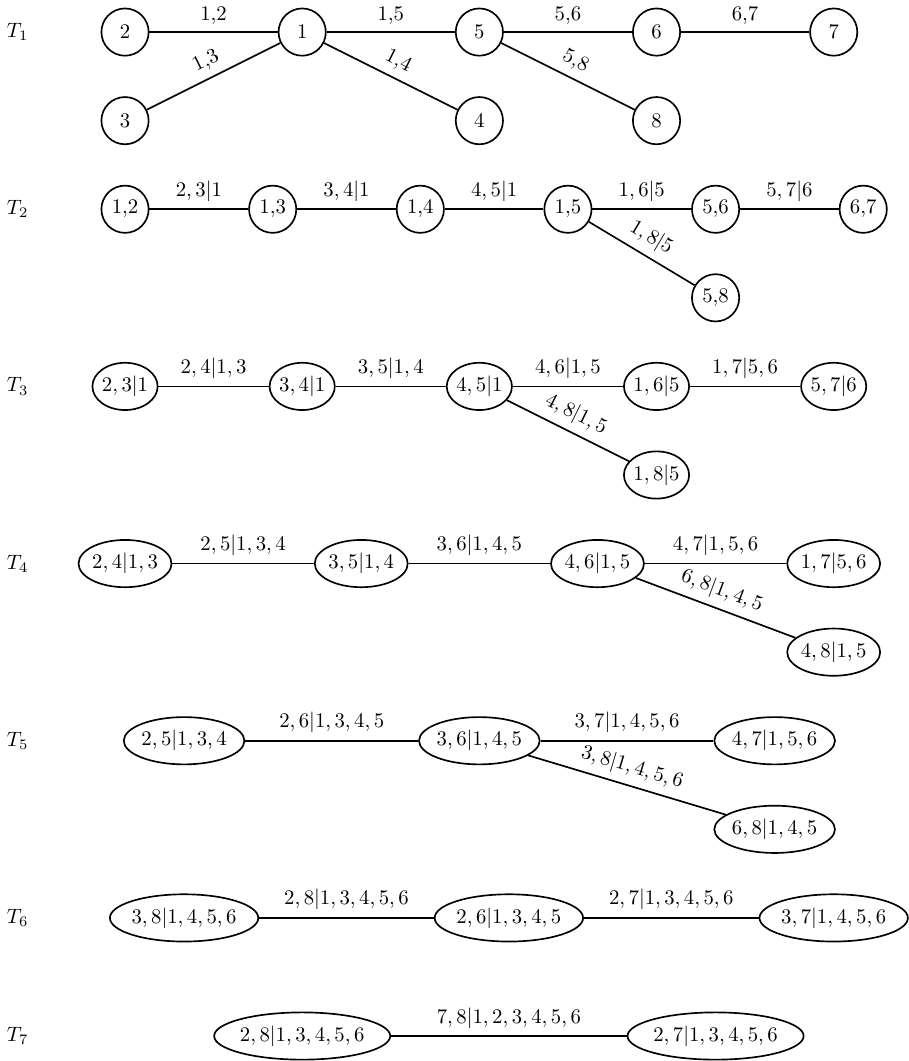}
	\caption{An R-vine tree sequence in 8 dimensions with edge indices corresponding to the pair-copulas in an R-vine copula model}
	\label{fig:neuer_Rvine}
\end{figure}

\noindent
\cite{MoralesNapoles2010} showed that there is a huge number of possible constructions. A set of nested trees is used to illustrate and order all these possible constructions. Each edge in a tree corresponds to a pair-copula in the PCC, while the nodes identify the pair-copula arguments.
\cite{Bedford_Cooke2001} formulated the following conditions, which a sequence of trees $\mathcal{V}=(T_1,\ldots,T_{d-1})$ has to fulfill to form an R-vine.
\begin{enumerate}
	\item $T_1$ is a tree with nodes $N_1=\{1,\ldots,d\}$ and edges $E_1$.
	\item For $i\geq 2$, $T_i$ is a tree with nodes $N_i = E_{i-1}$ and edges $E_i$.
	\item If two nodes in $T_{i+1}$ are joint by an edge, the corresponding edges in $T_i$ must share a common node {\it (proximity condition)}.
\end{enumerate}
\noindent
In our notation we follow \cite{DissmannBrechmannCzadoKurowicka2011} by denoting the vine structure with $\mathcal{V}$, the set of bivariate copulas with $\mathcal{B}(\mathcal{V})$ and the corresponding copula parameter with $\btheta(\mathcal{B}(\mathcal{V}))$. A specified regular vine copula we denote by $RV(\mathcal{V},\mathcal{B}(\mathcal{V}),\btheta(\mathcal{B}(\mathcal{V})))$.\\
The special cases of an R-vine tree structure $\mathcal{V}$ are line like and star structures of the trees. The first one is called D-vine in which each node has a maximum degree of 2, while the second one is a C-vine with a root node of degree $d-1$. All other nodes, so called leafs, have degree 1. D-vines are for example studied in \cite{Aas_Czado} or \cite{MinCzado2010}. An introduction to the statistical inference and model selection for C-vines are given for example in \cite{CzadoSchepsmeierMin2011}.\\
The pair-copula selection for an R-vine copula can be done by AIC, BIC or bivariate goodness-of-fit tests, while for the structure selection two algorithms are available in the literature. \cite{DissmannBrechmannCzadoKurowicka2011} favor a maximum spanning tree (MST) algorithm maximizing absolute Kendall's $\tau$ tree-wise, whereas \cite{GruberCzado2012} follow a Bayesian approach. They propose an algorithm to select the tree structure as well as the pair-copula families in a regular vine copula model jointly in a tree-by-tree reversible jump MCMC approach.
Figure \ref{fig:neuer_Rvine} gives the tree structure of an 8-dimensional R-vine fitted to our exchange rate data set used in Section \ref{sec:application} using the MST approach.\\ 

\section{The misspecification test problem}
\label{sec:test}

\noindent
Let $\boldsymbol{U}=(U_1,\ldots,U_d)^T\in[0,1]^d$ be a $d$ dimensional random vector with distribution function $G(\bu) = C_{\btheta}(u_1,\ldots,u_d)$, where $C_{\btheta}$ is a $d$ dimensional copula with parameter $\btheta$ and $U_i\sim\mathcal{U}(0,1)$ for $i=1,\ldots,d$. Let $c_{\btheta}$ be the corresponding copula density, then
\begin{gather}
\begin{split}
	\bbh(\btheta) &= E\left[ \partial_{\btheta}^2 ln(c_{\btheta}(U_1,\ldots,U_d))  \right], \\
	\bbc(\btheta) &= E\left[ \partial_{\btheta} ln(c_{\btheta}(U_1,\ldots,U_d)) \big( \partial_{\btheta} ln(c_{\btheta}(U_1,\ldots,U_d)) \big)^T \right]
\end{split}
\label{eq:HC}
\end{gather}
are the expected Hessian matrix of $ln(c_{\btheta}(u_1,\ldots,u_d))$ and the expected outer product of the corresponding score function, respectively, where $\partial_{\btheta}$ denotes the gradient with respect to the copula parameter $\btheta$. \\
Now, the theorem of \cite{White1982} shows that under correct model specification $(\btheta=\btheta_0)$ the negative expected Hessian matrix $\bbh(\btheta_0)$ and the expected outer product of the corresponding score function $\bbc(\btheta_0)$ are equal, i.e.
\begin{equation}
	-\bbh(\btheta_0) = \bbc(\btheta_0).
	\label{eq:equality}
\end{equation}
The corresponding vine copula misspecification test problem is therefore given by
\begin{equation}
	H_0: \bbh(\btheta_0) + \bbc(\btheta_0) = 0 \text{ against } H_1: \bbh(\btheta_0) + \bbc(\btheta_0) \neq 0,
	\label{eq:test}
\end{equation}
where $\btheta_0$ denotes the true value of the vine copula parameter vector.\\
In this paper we will expand the goodness-of-fit test of \cite{HuangProkhorov2011} who used White's information matrix equality (\ref{eq:equality}) to the vine copula.
In our setting we replace the $d$-dimensional density $c_{\btheta}(u_1,\ldots,u_d)$ in (\ref{eq:HC}) with the vine density $f_{1,\ldots,d}(u_1,\ldots,u_d)$ given in (\ref{eq:density}), with a parametric vine copula, i.e.~the parameter vector $\btheta$ is given by $\btheta=(\theta_{j(e),k(e);D(e)},e\in E_i, i=1,\ldots,d)$. In the following we assume that we have copula data available and therefore the vine density in (\ref{eq:density}) will be replaced by a vine copula density.

\section{Goodness-of-fit test for R-vine copulas}
\label{sec:gofRvine}

\noindent
The first step in the development of a goodness-of-fit test for the testing problem given in (\ref{eq:test}) will be the estimation of the Hessian matrix $\bbh(\btheta)$ and the outer product of gradient $\bbc(\btheta)$. For this we assume the availability of an i.i.d.~sample in $d$ dimensions of size $n$, denoted by $\bu:=(\bu_1,\ldots,\bu_n)$ in the copula space. This pseudo data $\bu$ is used to estimate the unknown true parameter $\btheta_0$ of the vine copula. Algorithms for maximum likelihood estimation for vine copula parameters are given in e.g.~\cite{CzadoSchepsmeierMin2011} for C-vine copulas or \cite{DissmannBrechmannCzadoKurowicka2011} for R-vine copulas. We call the estimate derived from these algorithms applied to $\bu_1,\ldots,\bu_n$ pseudo maximum likelihood (ML) estimate and denote it by $\hat{\btheta}_n=\hat{\btheta}(\bu_1,\ldots,\bu_n)$. 
Assuming that $\bU$ follows an R-vine copula model, i.e.~$\bU\sim RV(\mathcal{V},\mathcal{B}(\mathcal{V}),\btheta(\mathcal{B}(\mathcal{V})))$ we define the random matrices
\[
	\bbh(\btheta|\bU):=\frac{\partial^2}{\partial^2\btheta}l(\btheta|\bU)\qquad\text{and}\qquad \bbc(\btheta|\bU):=\frac{\partial}{\partial\btheta}l(\btheta|\bU)\left(\frac{\partial}{\partial\btheta}l(\btheta|\bU)\right)^T
\]
for the second derivative of the log-likelihood function $l(\btheta|\bU)$ and the outer product of the score function, respectively. Further, we denote the sample counter parts of the Hessian matrix and the outer product of the score function for the copula data $\bu_t=(u_{1t},\ldots,u_{dt})^T, 1\leq t \leq n$ at $\btheta=\hat{\btheta}_n$ of a vine copula with density given in (\ref{eq:density}) (in (\ref{eq:density}) the margins are assumed to be uniform) by
\[
	\hat{\bbh}_t(\hat{\btheta}_n):=\bbh(\hat{\btheta}_n|\bu_t)\in \bbr^{p\times p}\qquad\text{and}\qquad \hat{\bbc}_t(\hat{\btheta}_n):=\bbc(\hat{\btheta}_n|\bu_t)\in \bbr^{p\times p}.
\]
Here $p$ is the length of the parameter vector $\btheta$.
Thus, the sample equivalents to $\bbh(\btheta)$ (expected Hessian) and $\bbc(\btheta)$ (expected outer product of gradient) for the pseudo ML estimate $\hat{\btheta}_n$ are
\begin{equation*}
\begin{split}
	\bar{\bbh}(\hat{\btheta}_n) := \frac{1}{n}\sum_{t=1}^n \hat{\bbh}_t(\hat{\btheta}_n)\qquad\text{and}\qquad
	\bar{\bbc}(\hat{\btheta}_n) := \frac{1}{n}\sum_{t=1}^n \hat{\bbc}_t(\hat{\btheta}_n),
\end{split}
\end{equation*}
given $n$ observations. \cite{StoeberSchepsmeier2012} provide algorithms for the calculation of the gradient as well as of the Hessian matrix for R-vines. Alternatively numerical versions based on finite differences can be used as well, unless they are not precise enough.\\
Note that the matrices $\hat{\bbh}_t(\hat{\btheta}_n)$ and $\hat{\bbc}_t(\hat{\btheta}_n)$ are of size $d(d-1)/2 \times d(d-1)/2$ if all pair-copulas are one-parametric. In this case we have $p=d(d-1)/2$. For each two or higher parametrized bivariate copula in the vine the dimension of the matrices increases by 1 or the number of additional parameters. An example for a two-parameter bivariate copula is the bivariate Student's t-copula. But the dimension of the information matrices decrease if independence copulas are used in the pair-copula construction. This often appears in higher trees since the dependence usually decreases as number of trees increases. For higher dimensional vines a truncation, i.e.~setting all pair-copula families of higher order trees to the independence copula, may be helpful to reduce the number of parameters significantly \citep{BrechmannCzadoAas2012}.\\
To formulate the test statistic we vectorize the sum of the expected Hessian matrix $\bbh(\btheta_0)$ and the expected outer product of gradient $\bbc(\btheta_0)$. Therefore we define the random vector
\[
	\boldsymbol{d}(\btheta|\bU) := vech(\bbh(\btheta|\bU) + \bbc(\btheta|\bU)) \in \bbr^{\frac{p(p+1)}{2}}
\]
and its empirical version by
\[
	\hat{\boldsymbol{d}}_t(\hat{\btheta}_n):=\boldsymbol{d}(\hat{\btheta}_n|\bu_t) \qquad\text{and}\qquad  \bar{\boldsymbol{d}}(\hat{\btheta}_n) := \frac{1}{n}\sum_{t=1}^n \hat{\boldsymbol{d}}_t(\hat{\btheta}_n) \in \bbr^{\frac{p(p+1)}{2}}.
\]
Note that because of symmetry only the lower triangle (including the diagonal) of the matrices has to be vectorized.
Further, assuming existence of derivative and finite expectation we define the expected gradient matrix of the random vector $\boldsymbol{d}(\btheta|\bU)$ as
\begin{equation*}
\begin{split}
	\nabla D_{\btheta} &:= E\left[\partial_{\btheta_k}\boldsymbol{d}_l(\btheta|\bU) \right]_{l=1,\ldots,\frac{p(p+1)}{2},k=1,\ldots,p} \in \bbr^{\frac{p(p+1)}{2}\times p},\quad \text{and}\quad \\
	\widehat{\nabla D_{\btheta}} &:= \frac{1}{n} \sum_{t=1}^n\left[\partial_{\btheta_k}\hat{\boldsymbol{d}}_l(\hat{\btheta}_n|\bu_t)\right]_{l=1,\ldots,\frac{p(p+1)}{2},k=1,\ldots,p} \in \bbr^{\frac{p(p+1)}{2}\times p}
\end{split}
\end{equation*}
as its estimate. \cite[Appendix]{White1982} derived the corresponding asymptotic covariance matrix of $\sqrt{n}\bar{\boldsymbol{d}}(\hat{\btheta}_n)$, which is given by
\begin{align}
\label{eq:varianceMatrix}
	V_{\btheta_0} = \mathbbm{E}\bigg[&(\boldsymbol{d}(\btheta_0|\bU)-\nabla D_{\btheta_0}\bbh(\btheta_0)^{-1}\partial_{\btheta_0} l(\btheta_0|\bU)) \nonumber \\
	&\big(\boldsymbol{d}(\btheta_0|\bU)-\nabla D_{\btheta_0}\bbh(\btheta_0)^{-1}\partial_{\btheta_0} l(\btheta_0|\bU)\big)^T\bigg] 
\end{align}
In particular $\sqrt{n}\bar{\boldsymbol{d}}(\hat{\btheta}_n)\stackrel{d}{\longrightarrow} N(0,V_{\btheta_0})$, as $n\rightarrow\infty$.\\
The following proposition of Whites theorem is valid under the assumptions A1-A10 of \cite{White1982}. This assures that $l(\hat{\btheta}_n|\bu_t)$ is a continuous measurable function and its derivatives exist; A10 assumes that $V_{\btheta_0}$ is nonsingular.\\

\begin{proposition}
Under the correct vine copula specification and suitable regularity conditions (A1-A10 in \citealp{White1982}) the information matrix test statistic is defined as
\begin{equation}
	\calT_n = n\left(\bar{\boldsymbol{d}}(\hat{\btheta}_n)\right)^T\hat{V}_{\hat{\btheta}_n}^{-1} \bar{\boldsymbol{d}}(\hat{\btheta}_n),
	\label{eq:teststatistic}
\end{equation}
where $\hat{V}_{\hat{\btheta}_n}^{-1}$ is an consistent estimate for the inverse asymptotic covariance matrix $V_{\btheta_0}$.
It follows that $\calT_n$ is asymptotically $\chi^2_{p(p+1)/2}$ distributed.\\
\end{proposition}

\noindent
The proof is an extension of the proof of \cite{White1982} since the maximum likelihood estimator $\hat{\btheta}_n$ in a vine copula is also to be shown normally distributed \citep{haff2010}.
\begin{equation}
\sqrt{n}\ \cali(\boldsymbol \theta_0)^{1/2}\Big(\hat{\boldsymbol \theta}_n - \boldsymbol \theta_0\Big) \stackrel{d}{\longrightarrow} N(0,\boldsymbol{Id}_p)\ \text{as}\ n\rightarrow\infty,
\label{eq:asymptotic_MLE}
\end{equation}
where, $\boldsymbol \theta_0$ is the true $p$-dimensional parameter vector, $n$ is the number of (i.i.d.) observations,  $\boldsymbol{Id}_p$ the $p\times p$ identity matrix, and $\cali(\boldsymbol \theta_0) = -\bbh(\btheta_0) = \bbc(\btheta_0)$.
Since the asymptotic distribution is independent of model parameters the test is asymptotically pivotal. 
The $\chi^2$-distribution only depends on the parameter vector dimension $p$, which is known aforehand given the pair-copula families, and not on $\btheta_0$. 
Furthermore, the test is a so called "blanket" test in the sense of \cite{Genest2009}.\\
 
\noindent
Note that all calculations are performed using copula data, thus ignoring the uncertainty in the margins. In Section \ref{sec:extension} we extend our goodness-of-fit test adjusting $V_{\btheta_0}$ for the estimation of the margins. For general multivariate copulas this is already done by \cite{HuangProkhorov2011}.\\
Given the test statistic $\calT_n$ of Proposition 1 we can define an $\alpha$-level test.\\

\begin{corollary}
Let $\alpha\in(0,1)$ and $\calT_n$ as in Proposition 1. Then the test
\begin{equation*}
\begin{split}
\text{Reject } &H_0: \bbh(\btheta_0) + \bbc(\btheta_0) = 0 \text{\quad versus\quad } H_1: \bbh(\btheta_0) + \bbc(\btheta_0) \neq 0 \\
&\Leftrightarrow \calT > \left(\chi^2_{p(p+1)/2}\right)^{-1}(1-\alpha)  
\end{split}
\end{equation*}
is an asymptotic $\alpha$-level test. Here $\left(\chi^2_{df}\right)^{-1}(\beta)$ denotes the $\beta$ quantile of a $\chi^2_{df}$ distribution with $df$ degrees of freedom.\\
\end{corollary}

\section{Power studies}
\label{sec:powerstudy}

\noindent
A goodness-of-fit test's performance is usually measured by its power, which is often unknown. In this case, it can only be investigated via simulations. Given a significance level $\alpha$ a high power at a specified alternative indicates a good discrimination against the alternative. In this section we will investigate the introduced test for a suitable large $n$ and 
under a variety of alternatives. 
In particular we determine the power function under the true data generating process (DGP) and under the alternative DGP. In the first case the null hypothesis $H_0$ holds, while in the second case the null hypothesis $H_0$ does not hold. The power at the true DGP assesses the ability of the test in Corollary 1 to maintain its nominal level.

\subsection{Performance measures}

\noindent
As performance measures we use p-value plots and size-power curves, introduced by \cite{DavidsonMacKinnon1998} and explained in the following.\\
Given an observed test statistic $\calT_n=t$ the p-value at $t$ is 
\[
	p(t):=P(\calT_n \geq t),
\]
i.e.~the smallest $\alpha$ level at which the test can reject $H_0$ when $\calT_n=t$ is observed. The distribution of $\calT_n$ has to be known at least asymptotically such as the $\chi^2$-distribution in our case or has to be estimated empirically using simulated data. \\
Let $Z_{M_1}:=p(\calT_n(M_1))$ be the random variable on $[0,1]$ with value $p(t)$ when $\calT_n=t$ is observed and data is generated from model $M_1$. The distribution function for $Z_{M_1}$ we denote by $F_{M_1}(\cdot)$, i.e.~$F_{M_1}(\alpha):=P(Z_{M_1}\leq\alpha)$, being the size of the test. To estimate $Z_{M_1}$ and $F_{M_1}$ we assume $R$ realizations of the test statistic $\calT_n(M_1)$ when $n$ observations are generated from model $M_1$, denoted as $\calt^j_n({M_1}), j=1,\ldots,R$, and estimate the p-values $p^j_{M_1}$ as
\[
	\hat{p}^j_{M_1}:=\hat{p}(\calt^j_n(M_1)) := \frac{1}{R}\sum_{r=1}^R \boldsymbol{1}_{\{\calt^r_n(M_1)\geq \calt^j_n(M_1)\}}
\]
and consider the empirical distribution function of them. Thus,
\[
	\hat{F}_{M_1}(\alpha) := \frac{1}{R}\sum_{r=1}^R \boldsymbol{1}_{\{\hat{p}^r_{M_1}\leq \alpha\}}\quad \alpha\in(0,1),
\]
forms an estimate for the size of the test at level $\alpha$. $\hat{F}_{M_1}(\alpha)$ is called \textit{actual} size or \textit{actual} alpha (probability of the outcome under the null hypothesis), while $\alpha$ is known as \textit{nominal} size.\\

\noindent
\textbf{p-value plot}\\
The p-value plot graphs $\alpha$ versus $\hat{F}_{M_1}(\alpha)$, i.e.~\textit{nominal} size against \textit{actual} size. The plot indicates if the test reaches its \textit{nominal} size, i.e.~if the assumed asymptotic holds.\\

\noindent
\textbf{Size-power curve}\\
We are not only interested in the size of the test but also in its power. For data generated under model $M_2$ we want to determine $F_{{M_2}}(\alpha)$, which gives the power of the test when $H_1$ is true, i.e.~$M_2$ holds, and a level $\alpha$ is used for the test. 
Generate $R$ i.i.d.~data sets from $M_2$ and use them to estimate $F_{{M_2}}(\alpha)$ by $\hat{F}_{{M_2}}(\alpha)$.
The plot of $\hat{F}_{M_1}(\alpha)$ versus $\hat{F}_{{M_2}}(\alpha)$ is called the size-power curve. 
A test with good power should have small power when the size is small and approach power one rapidly as the size increases.\\

\noindent
Remark: Note that these curves correspond to the better known Receiver-operating-characteristic curves (ROC), which plot the fraction of true positives (TP) out of the positives (TPR = true positive rate) versus the fraction of false positives (FP) out of the negatives (FPR = false positive rate), at various threshold settings. TPR is also known as sensitivity, and FPR is one minus the specificity or true negative rate. False positive (FP) is better known as Type I error. Since $\alpha$ (\textit{nominal} size) is the false positive rate and power is one minus the false negative rate (FNR; prob. of of a Type II error occurring), ROC plots \textit{nominal} size versus power \citep{Fawcett2006}.

\subsection{General remarks on size and power, and the implementation of the test}

\begin{itemize}
\item It is shown for other statistical models that the size behavior of the information matrix test (IMT) is very poor. E.g.~in the regression model context the IMT has poor size properties even in samples of size 1000 and more (see \cite{Hall1989} and references within, especially \cite{Taylor1987}). Similar observations are made by e.g.~\cite{ChesherSpady1991}.
	\item In the bivariate copula case the asymptotic approximation holds even for relative small number of observations. But this is not investigated or documented in \cite{HuangProkhorov2011}.
	\item In this simulation study three possible errors can occur, which may influence the asymptotic behavior: simulation error, estimation and model error, and numerical errors. Simulation errors are always involved since only pseudo random variables can be generated on a computer. For the parameter estimation maximum likelihood is used based on Newton-Raphson algorithms for maximization. In higher dimensions this can be quite challenging, even given the analytical gradient (and the analytical Hessian matrix). A local maximum may be returned. Further, numerical instabilities can occur, especially in the calculation of the score function and the Hessian matrix as discussed in \cite{StoeberSchepsmeier2012}.
	\item Even the estimator of $V_{\btheta_0}$ may not be positive definite, though this becomes increasingly unlikely as the sample size increases.
	\item Furthermore, the normal asymptotic theory only holds for full maximum likelihood, but a sequential maximum likelihood, i.e.~a tree-wise estimation, is performed due to resource and time limits. Usually sequential estimates are close to full ML estimates, except of the degree-of-freedom parameter $\nu$ of the Student's t copula. There exists even a asymptotic theory for sequential estimates similar to Equation (\ref{eq:asymptotic_MLE}), see \cite{haff2010}.
\end{itemize}

\subsection{General simulation setup}

\noindent
We test if our goodness-of-fit test $\calT_n$ based on the vine copula model $M_1=RV(\mathcal{V}_1,\mathcal{B}_1(\mathcal{V}_1),\btheta_1(\mathcal{B}_1(\mathcal{V}_1)))$ has suitable power against an alternative vine copula model $M_2=RV(\mathcal{V}_2,\mathcal{B}_2(\mathcal{V}_2),\btheta_2(\mathcal{B}_2(\mathcal{V}_2)))$, where $M_2\neq M_1$. To produce the corresponding p-value plots for $M_1$ and the size-power curves 
 we proceed as follows:
\begin{enumerate}
	\item Set vine copula model $M_1$.
	\item Generate a copula data sample of size $n=1000$ from model $M_1$ (pre-run).
	\item Given the data of the pre-run select and estimate $M_2$ using e.g.~the step-wise selection algorithm of \cite{DissmannBrechmannCzadoKurowicka2011}.
	\item For $r=1,\ldots,R$
	\begin{itemize}
		\item Generate copula data $\boldsymbol{u}_{M_1}^r=(\boldsymbol{u}_{M_1}^{1r},\ldots,\boldsymbol{u}_{M_1}^{dr})$ from $M_1$ of size $n$. 
		\item Estimate $\btheta_1(\mathcal{B}_1(\mathcal{V}_1))$ of model $M_1$ given data $\boldsymbol{u}_{M_1}^r$ and denote it by $\hat{\btheta}_1(\mathcal{B}_1(\mathcal{V}_1);\bu_{M_1}^r)$.
		\item Calculate test statistic $t_n^r(M_1):=t_n^r(\hat{\btheta}_1(\mathcal{B}_1(\mathcal{V}_1);\bu_{M_1}^r))$ based on data $\boldsymbol{u}_{M_1}^r$ assuming the vine copula model $M_1=RV(\mathcal{V}_1,\mathcal{B}_1(\mathcal{V}_1),\hat{\btheta}_1(\mathcal{B}_1(\mathcal{V}_1)))$.
		\item Calculate asymptotic p-values $p(t_n^r(M_1))=(\chi^2_{p(p+1)/2})^{-1}(t_n^r(M_1))$, \\ 
	where $p$ is the number of parameters of $\btheta_1(\mathcal{B}_1(\mathcal{V}_1))$.
		
	\item Generate copula data $\boldsymbol{u}_{M_2}^r=(\boldsymbol{u}_{M_2}^{1r},\ldots,\boldsymbol{u}_{M_2}^{dr})$ from $M_2$ of size $n$.  
	\item Estimate $\btheta_1(\mathcal{B}_1(\mathcal{V}_1))$ of model $M_1$ given data $\boldsymbol{u}_{M_2}^r$ and denote it by $\hat{\btheta}_1(\mathcal{B}_1(\mathcal{V}_1);\bu_{M_2})$. 
	\item Calculate test statistic $t_n^r(M_2)$ based on data $\boldsymbol{u}_{M_2}^r$ assuming vine copula model $M_1$.
	\item Calculate asymptotic p-values $p(t_n^r(M_2))=(\chi^2_{p(p+1)/2})^{-1}(t_n^r(M_2))$. \\ 
	\end{itemize}
	end for
	\item Estimate p-values $p^j_{M_1}$ and $p^j_{M_2}$ by \\
	\begin{equation*}
	\begin{split}
	\hat{p}^j_{M_1} = \hat{p}(\calt^j_n(M_1)) &:= \frac{1}{R}\sum_{r=1}^R \boldsymbol{1}_{\{\calt^r_n(M_1)\geq \calt^j_n(M_1)\}}\quad\text{and} \\
	\hat{p}^j_{M_2} = \hat{p}(\calt^j_n(M_2)) &:= \frac{1}{R}\sum_{r=1}^R \boldsymbol{1}_{\{\calt^r_n(M_2)\geq \calt^j_n(M_2)\}}, 
	\end{split}
	\end{equation*}
	 respectively, for $j=1,\ldots,R$.
	\item Estimate the distribution function of $Z_{M_1}$ and $Z_{M_2}$ by  
	\[ 
		\hat{F}_{M_1}(\alpha) = \frac{1}{R}\sum_{r=1}^R \boldsymbol{1}_{\{\hat{p}^r_{M_1}\leq \alpha\}} \quad\text{and}\quad
		\hat{F}_{M_2}(\alpha) = \frac{1}{R}\sum_{r=1}^R \boldsymbol{1}_{\{\hat{p}^r_{M_2}\leq \alpha\}},
	\]
	respectively.
\end{enumerate}

\noindent
The following simulation results are based on $R=10000$ replications and the number of observations $n$ are chosen to be 300, 500, 750 or 1000. The dimension of the vine copula models $M_i, i=1,2$ is 5. Possible pair-copula families are the elliptical Gauss and Student's t-copula, the Archimedean Clayton, Gumbel, Frank and Joe copula, and the rotated Archimedean copulas.
A p-value plot to assess the nominal size of the test is achieved by plotting $\alpha$ versus $\hat{F}_{M_1}(\alpha)$.
Evaluating $\hat{F}_{M_1}(\alpha)$ and $\hat{F}_{M_2}(\alpha)$ on the grid 
\[
\alpha=0.001, 0.002, \ldots, 0.010, 0.015, \ldots,0.990, 0.991, \ldots, 0.999
\]
with smaller grid size near 0 and 1 we can plot a size-power curve $\hat{F}_{M_1}(\alpha)$ versus $\hat{F}_{M_2}(\alpha)$.\\
All calculations are performed with R\footnote{R Development Core Team (2012). R: A language and environment for statistical computing. R Foundation for Statistical Computing, Vienna, Austria. ISBN 3-900051-07-0, URL http://www.R-project.org/.}, the R-package \texttt{VineCopula} of \cite{VineCopula} and the \texttt{copula}-package of \cite{copula}.\\

\subsection{Specific setting}

\noindent
In the following three power studies we investigate the properties of the introduced test with respect to its size and power. 
In the first power study we determine the power of the test assuming an R-vine copula as true model ($M_1$ in the notation from above) under three alternatives of simpler copula models such as 
\begin{itemize}
	\item the multivariate Gauss copula,
	\item the C-vine copula and
	\item the D-vine copula,
\end{itemize}
which are special cases of the R-vine.
Every multivariate Gaussian copula can be written as a vine copula with Gaussian pair-copulas and vice versa \cite{Czado}. Only in the Gaussian case the conditional correlation parameters, forming the pair-copula parameters, are equal to the partial correlation parameter, which can be calculated recursively using the entries of the multivariate Gauss copula variance-covariance matrix.\\
The second power study investigates the power of the test between two R-vines, which are chosen with two different selection methods - a maximum spanning tree approach introduced by \cite{DissmannBrechmannCzadoKurowicka2011} versus a Bayesian approach (MCMC) investigated by \cite{GruberCzado2012}, based on a generated data set given a specified R-vine copula model.\\
In a third simulation study we compare the often used multivariate t-copula under the alternative of an R-vine with only bivariate t-copulas, and vise versa. The difference is the common degree-of-freedom parameter $\nu$ in the multivariate t-copula versus variable, separately estimated $\nu$s in the R-vine model. The correlation parameters $\rho$ can be set/estimated such as in the Gaussian case described above.\\
Table \ref{tab:OverviewOfTheStudiedTestSettings} gives an overview of all three power studies, their true model and their alternatives.

\begin{table}[htbp]
	\centering
	{\footnotesize
		\begin{tabular}{lll}
			\toprule
			Study & True model ($M_1$) & Alternative ($M_2$) \\
			\midrule
			I & R-vine  & {\footnotesize multivariate Gauss} \\
				& $(\mathcal{V}_{R},\mathcal{B}_{R}(\mathcal{V}_{R}),\btheta_{R}(\mathcal{B}_{R}(\mathcal{V}_{R})))$ & C-vine $(\mathcal{V}_{C},\mathcal{B}_{C}(\mathcal{V}_{C}),\btheta_{C}(\mathcal{B}_{C}(\mathcal{V}_{C})))$ \\
				& & D-vine $(\mathcal{V}_{D},\mathcal{B}_{D}(\mathcal{V}_{D}),\btheta_{D}(\mathcal{B}_{D}(\mathcal{V}_{D})))$ \\
			II & R-vine  & R-vine estimated by \cite{DissmannBrechmannCzadoKurowicka2011} \\
				& $(\mathcal{V}_{R},\mathcal{B}_{R}(\mathcal{V}_{R}),\btheta_{R}(\mathcal{B}_{R}(\mathcal{V}_{R})))$ & $(\mathcal{V}_{MST},\mathcal{B}_{MST}(\mathcal{V}_{MST}),\btheta_{MST}(\mathcal{B}_{MST}(\mathcal{V}_{MST})))$ \\
				& & R-vine estimated by \cite{GruberCzado2012} \\
				& & $(\mathcal{V}_{B},\mathcal{B}_{B}(\mathcal{V}_{B}),\btheta_{B}(\mathcal{B}_{B}(\mathcal{V}_{B})))$ \\
			III & multivariate t-copula & R-vine estimated by \cite{DissmannBrechmannCzadoKurowicka2011} \\
				& & with only Student's t-copulas \\
				& R-vine with only t-copulas & multivariate t-copula \\
		\bottomrule
		\end{tabular}}
	\caption{Overview of the studied test settings}
	\label{tab:OverviewOfTheStudiedTestSettings}
\end{table}

%



\subsubsection{Power study I}
\noindent
We investigated three variants of the dependence: 
\begin{itemize}
	\item $M_1$ with mixed Kendall's $\tau$ values,
	\item $M_1$ with constant low ($\tau=0.25$) Kendall's $\tau$ values and 
	\item $M_1$ with constant medium ($\tau=0.5$) Kendall's $\tau$ values
\end{itemize}
for the $d(d-1)/2$ pair-copulas. An R-vine with constant high dependencies is omitted since the power in the medium case are already very high and allow to draw conclusions for the high dependency case. The structure of the chosen R-vine is given in Figure \ref{fig:5dimRvine} and Equation (\ref{eq:5dimRvine}) of \ref{appendix:powerstudy}. The chosen bivariate copula families and Kendall's $\tau$ values can be found in Table \ref{tab:5dimRvine} of \ref{appendix:powerstudy}.\\
The selected D-vine structure (Step 3 in the test procedure) is already defined by the ordering of its variables in the first tree. Here the ordering is 3-4-5-1-2, see Equation (\ref{eq:5dimDvine}) of \ref{appendix:powerstudy}. Similarly, the C-vine structure is defined by its root nodes. The root in the first tree is variable 2 while in the second tree variable 1 is added to the root, i.e.~the root in Tree 2 is 1,2. Variable 4, 5 and 3 are added in the next trees, respectively, see Equation (\ref{eq:5dimCvine}) of  \ref{appendix:powerstudy}. The selected copula structure and pair-copula parameters in Step 3 are quite stable given more than one data set in Step 2, i.e.~no changes in the vine copula structure and minor changes in the pair-copula choice (e.g.~the algorithm selects a rotated Gumbel copula instead of a Clayton copula, which are quite similar given a low Kendall's $\tau$). Neglecting possible small variations in the pair-copula selection given $R$ different data sets we fix the C- and D-vine structure as well as the pair-copula family selection after one run of Step 2.
An overview of all investigated models is given in Table \ref{tab:overviewModels}.

\begin{table}[htbp]
	\centering
	{\footnotesize
		\begin{tabular}{lccccc}
			\toprule
			Model & $\mathcal{V}$ & $\mathcal{B}(\mathcal{V})$ & $\btheta(\mathcal{B}(\mathcal{V}))_1$ & $\btheta(\mathcal{B}(\mathcal{V}))_2$ & $\btheta(\mathcal{B}(\mathcal{V}))_3$\\
			\midrule
			R-vine & $\hat{\mathcal{V}}_R$, Eq. (\ref{eq:5dimRvine}) & Tab. \ref{tab:5dimRvine} & Tab. \ref{tab:5dimRvine} & $\tau=0.25$ & $\tau=0.5$ \\
			Gauss & - & Gauss & $\hat{\btheta}_1$ & $\hat{\btheta}_2$ & $\hat{\btheta}_3$ \\
			C-vine & $\hat{\mathcal{V}}_C$, Eq. (\ref{eq:5dimCvine}) & $\hat{\mathcal{B}}_C(\hat{\mathcal{V}}_C)$ & $\hat{\btheta}_C(\hat{\mathcal{B}}_C(\hat{\mathcal{V}}_C))_1$ & $\hat{\btheta}_C(\hat{\mathcal{B}_C}(\hat{\mathcal{V}}_C))_2$ & $\hat{\btheta}_C(\hat{\mathcal{B}}_C(\hat{\mathcal{V}}_C))_3$ \\
			D-vine & $\hat{\mathcal{V}}_D$, Eq. (\ref{eq:5dimDvine}) & $\hat{\mathcal{B}}_D(\hat{\mathcal{V}}_D)$ & $\hat{\btheta}_D(\hat{\mathcal{B}}_D(\hat{\mathcal{V}}_D))_1$ & $\hat{\btheta}_D(\hat{\mathcal{B}}_D(\hat{\mathcal{V}}_D))_2$ & $\hat{\btheta}_D(\hat{\mathcal{B}}_D(\hat{\mathcal{V}}_D))_3$ \\
			
			\bottomrule
		\end{tabular}}
	\caption{Model specifications}
	\label{tab:overviewModels}
\end{table}

\begin{figure}[!ht]
	\centering
	\subfigure[p-value plots using simulated p-values]{\includegraphics[width=0.90\textwidth]{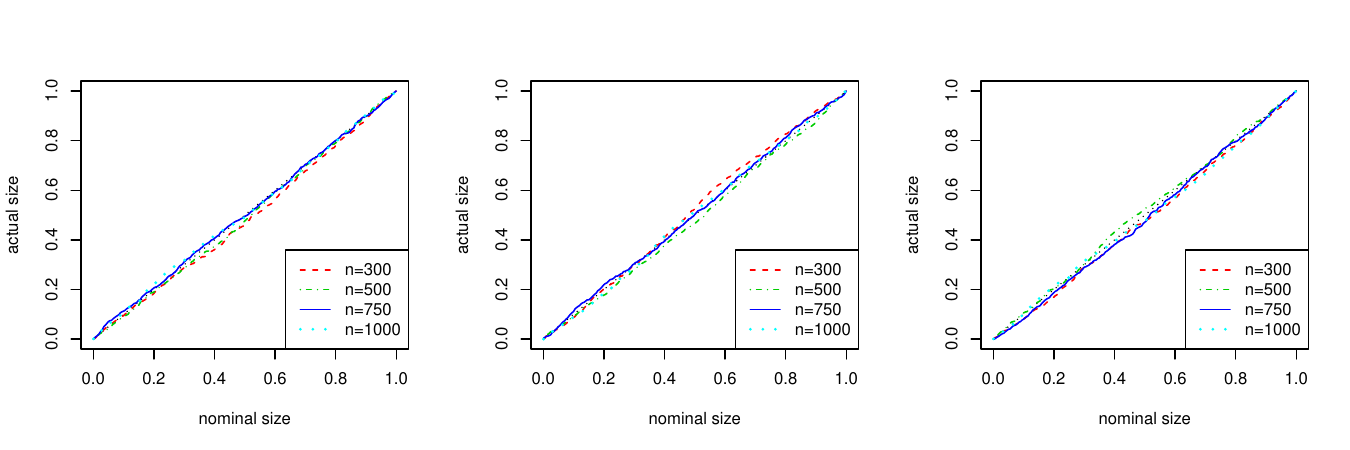}}
	\subfigure[p-value plots using asymptotic p-values]{\includegraphics[width=0.90\textwidth]{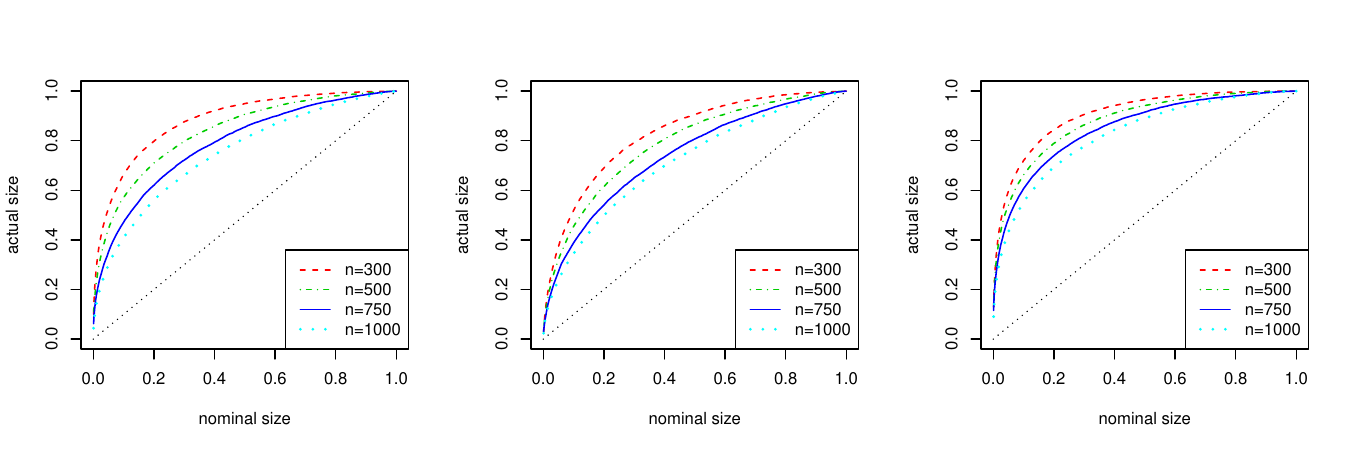}}
	\subfigure[detail plots of (b) ]{\includegraphics[width=0.95\textwidth]{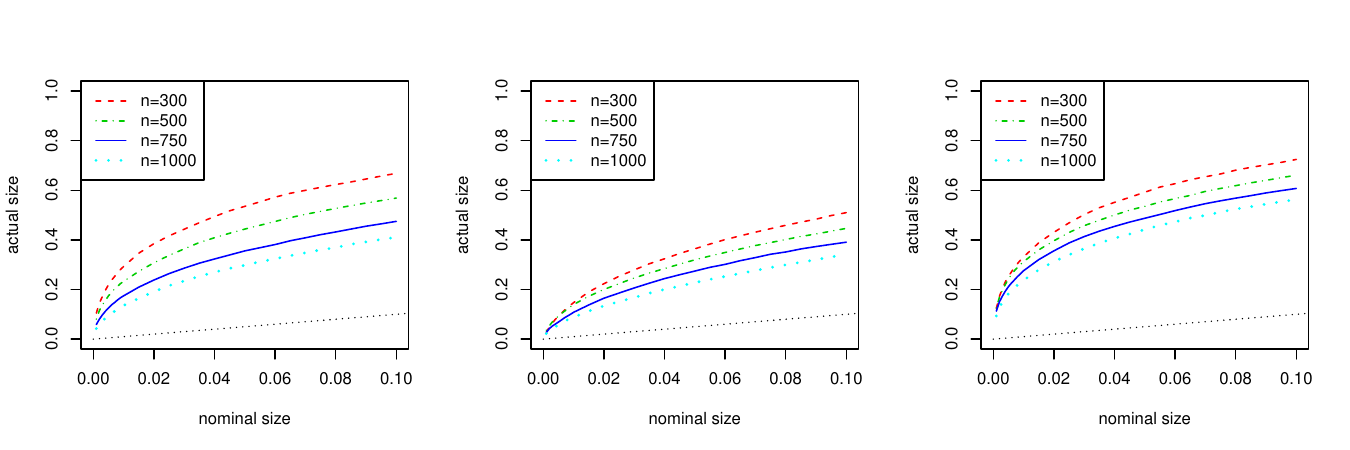}}
	\caption{p-value plots for the three different scenarios; left: mixed Kendall's $\tau$, center: constant Kendall's $\tau=0.25$, right: constant Kendall's $\tau=0.5$.}
	\label{fig:pvaluePlotAsy}
\end{figure}

\begin{figure}[!ht]
\centering
		\subfigure[$d=5$]{\includegraphics[width=0.90\textwidth]{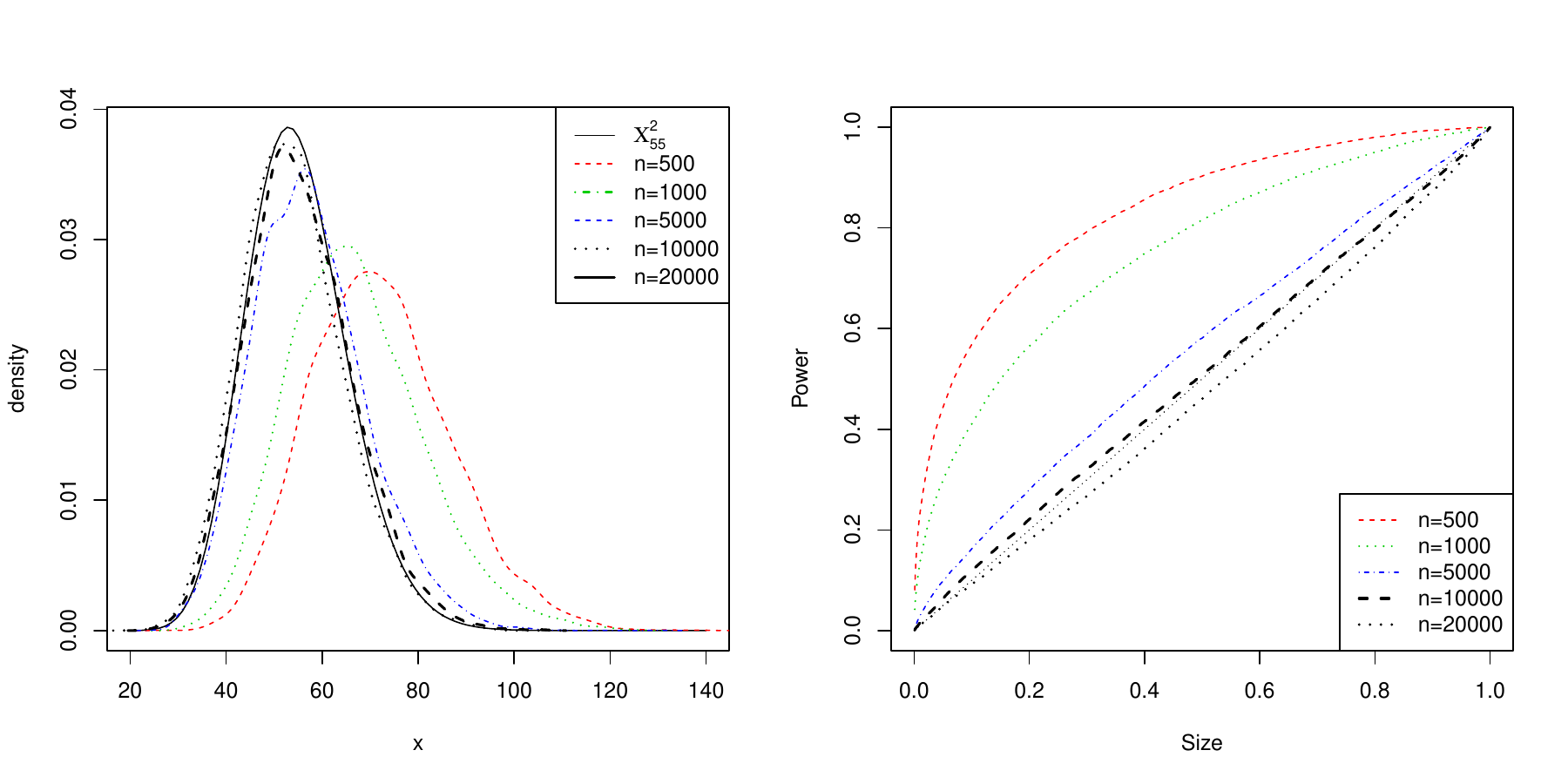}}
	\subfigure[$d=8$]{\includegraphics[width=0.90\textwidth]{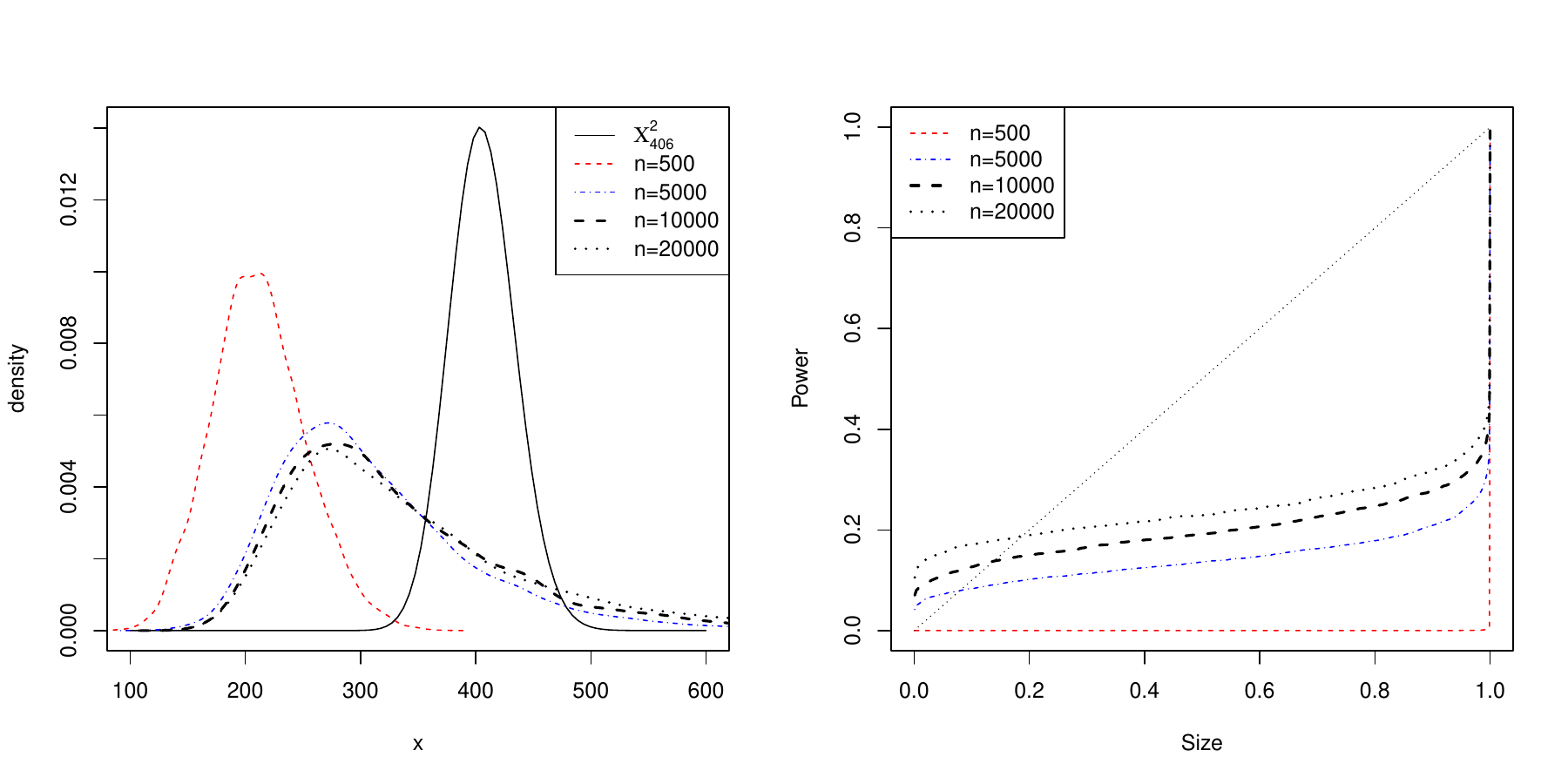}}
	\caption{Empirical density plot (left panel) and p-value plot for the asymptotic p-values (right panel)}
	\label{fig:pvalueHighN}
\end{figure}

\noindent
\textbf{Results with regard to nominal size}\\
A p-value plot ($\alpha$ versus $\hat{F}_{M_1}(\alpha)$) for the simulated results shows that the test works perfectly under the null independent of the number of observations, since the p-value plot fits the 45 degree line nearly perfect (see Figure \ref{fig:pvaluePlotAsy}a). That means that the test reaches its nominal level in the case of simulated p-values.\\
Given the asymptotic p-values $p(t_n^r(M_1))=(\chi^2_{p(p+1)/2})^{-1}(t_n^r(M_1))$ and their corresponding empirical distribution function $\hat{F}_{M_1}^{asy}(\alpha)$ we have a different picture. The \textit{actual} size is much greater than the \textit{nominal} size (see Figure \ref{fig:pvaluePlotAsy}b). Also the test does not hold its nominal level in the case of asymptotic p-values given a small/medium data set. The test over-rejects quite too often based on asymptotic p-values. \\
Comparing the finite sample distribution of the test statistic with the theoretical $\chi^2$ distribution in Figure \ref{fig:pvalueHighN}a (left panel) we can clearly see, that even in the case of 1000 observation points the $\chi^2$ distribution does not fit the empirical distribution given the observed test statistics $t_n^r(M_1)$. For the investigated 5-dimensional case the empirical distribution fits the theoretical one not until $n=10000$ . In this case the actual size is the nominal size (see Figure \ref{fig:pvalueHighN}a, right panel).\\
Additionally, we investigated the size behavior in a 8-dimensional vine copula model, whose details are not provided here. Figure \ref{fig:pvalueHighN}b clearly illustrates that the asymptotic theory based test is too conservative, while in the 5-dimensional case the test is too liberal. Even for a sample size of $n=20000$ for $d=8$ the the actual size does not reach the nominal size.\\

\noindent

\begin{figure}[htbp]
	\centering
		\subfigure[empirical with mixed Kendall's $\tau$]{\includegraphics[width=0.47\textwidth]{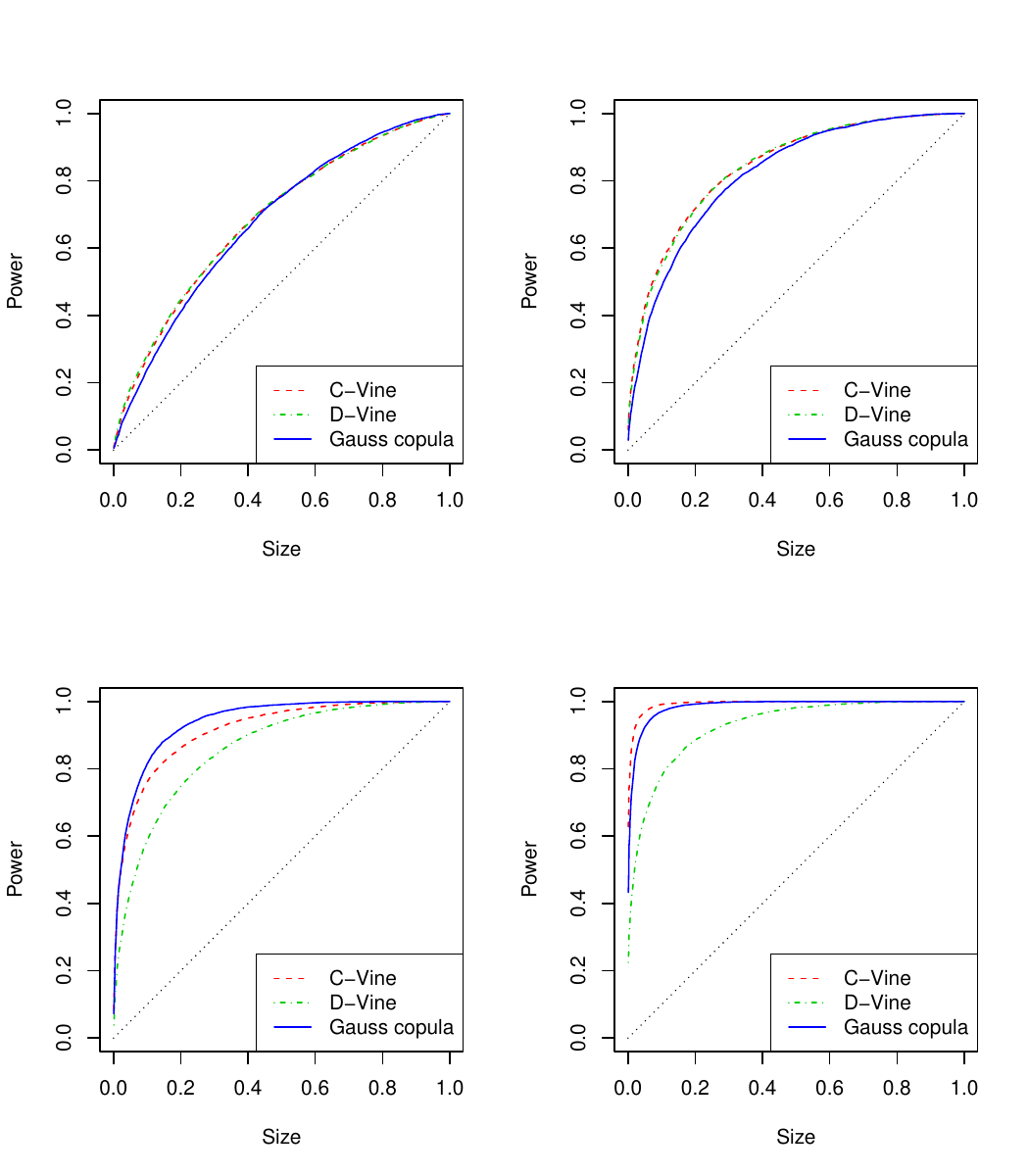}}  
		\subfigure[empirical with constant Kendall's $\tau=0.5$]{\includegraphics[width=0.47\textwidth]{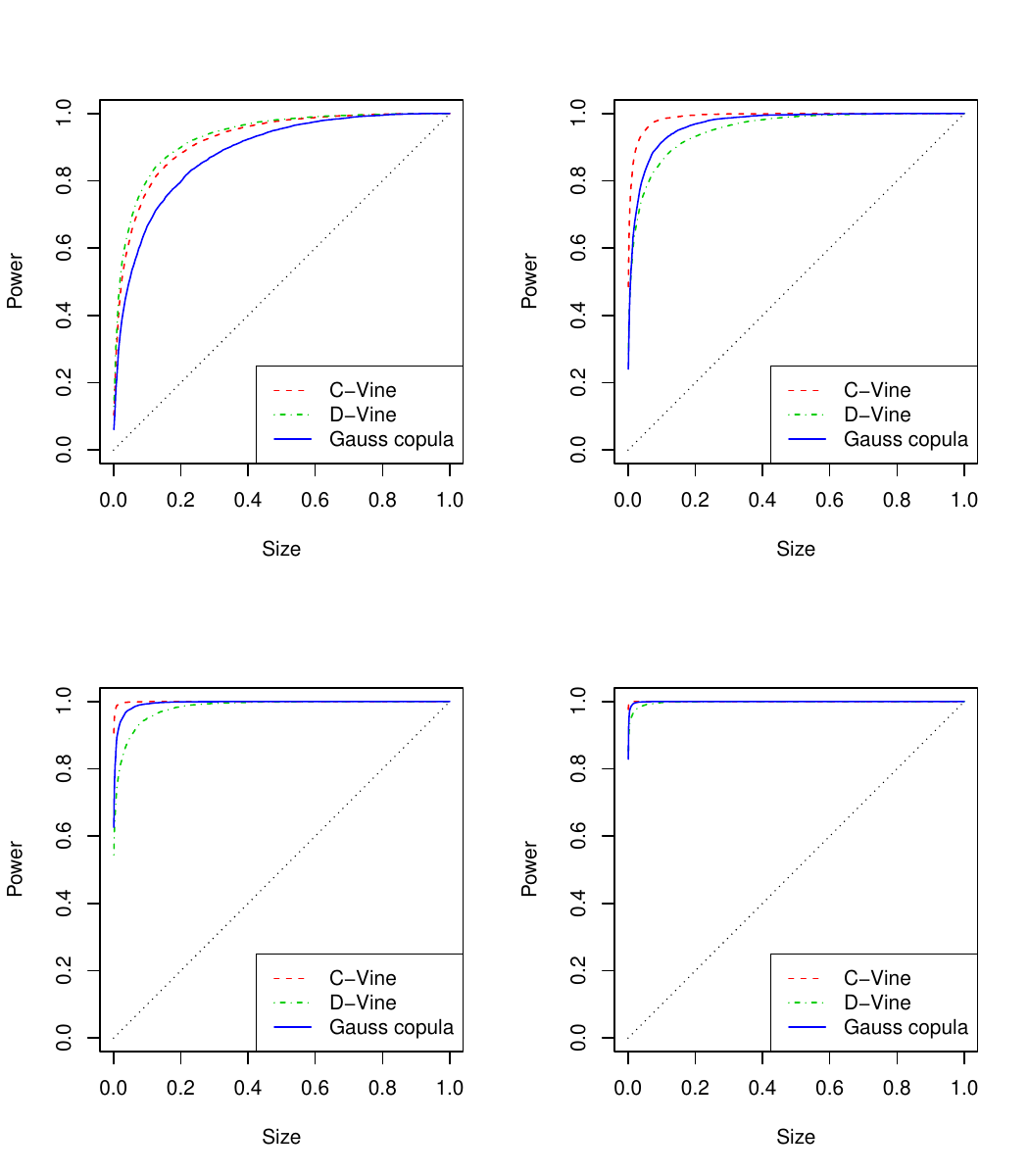}} \hfill 
		\subfigure[empirical with constant Kendall's $\tau=0.25$]{\includegraphics[width=0.47\textwidth]{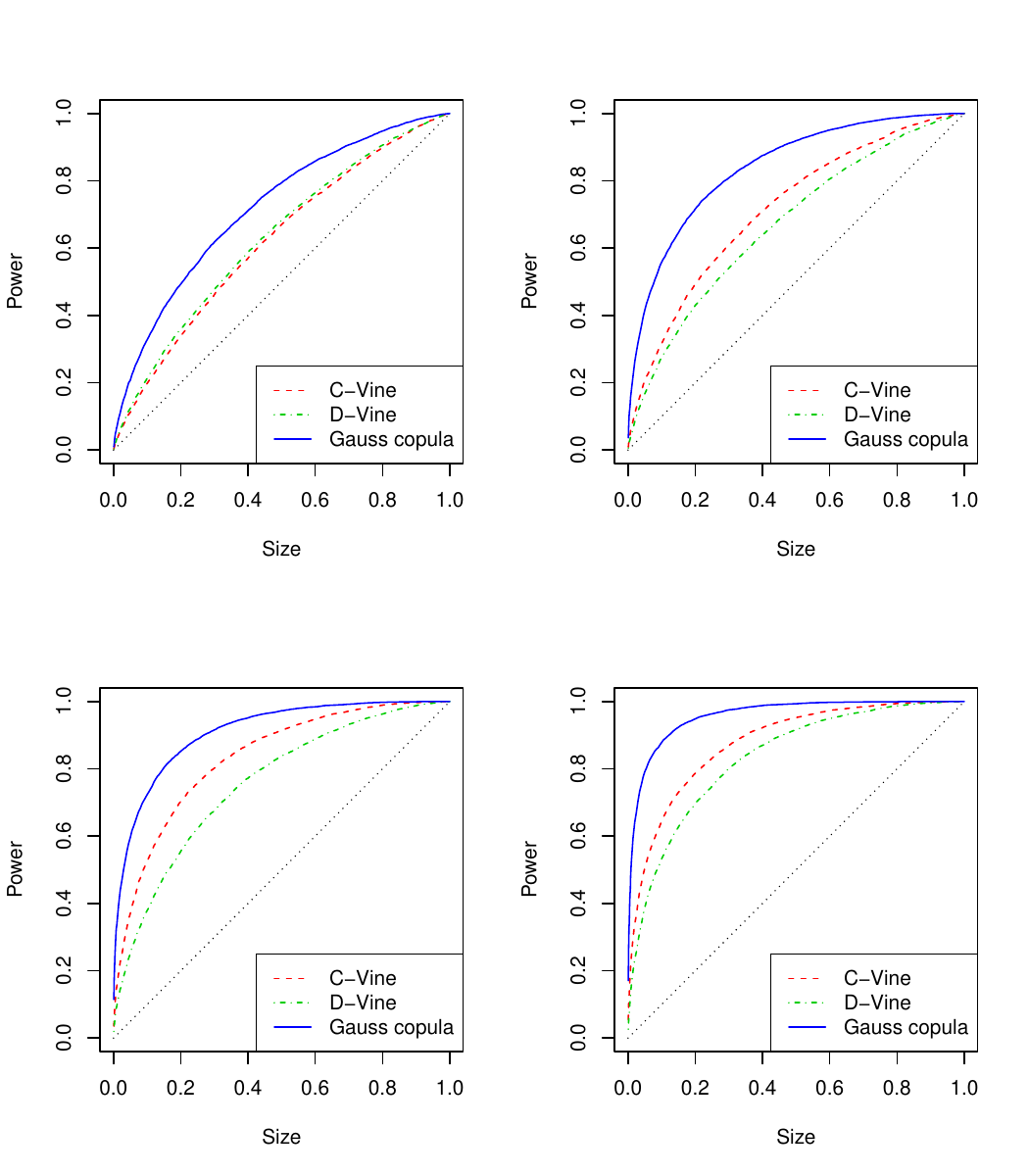}} 
	\caption{Size-power curves for different number of observations. 
	In each panel: Upper left: $n=300$, upper right: $n=500$, lower left: $n=750$ and lower right: $n=1000$.}
	\label{fig:powerstudy2}
\end{figure}

\noindent
\textbf{Results with regard to power}\\
In Figure \ref{fig:powerstudy2} we show the behavior of the size-power curve for varying number of observations $n$ in each scenario. The number of observations increase from $n=300$ in the upper left panel over $n=500$ (upper right panel) and $n=750$ (lower left panel) to $n=1000$ in the lower right panel. Due to the results of the p-value plots we only consider the results of the simulated p-values in the following. The dotted diagonal line represents the case where size (x-axes) equals power (y-axes). 
In addition, we list in Table \ref{tab:powerTable} the power for $n=500$ at \textit{nominal} size 5\% in several senarios. Given a true vine model different vine models are tested. E.g.~given a C-vine with Kendall's $\tau$ value of 0.25 the test for an R-vine returned on average a power of 18.4\% using simulated p-values (Simul.) and 59.9\% using asymptotic p-values (Asy.).\\

\noindent
The first observation evaluating the plots is that the power is always greater than the size. This indicates a good performance of the test in mean. Further, an increasing number of observations increases the power of the test. In the medium dependence scenarios the tests are consistent since the power reaches one at quite low size.

\noindent
Further conclusions are:
\begin{itemize}
	\item The size-power curves of the C-vine and the D-vine are close to each other in each scenario. This changes in higher dimensional data sets (a 8-dimensional scenario was performed but is not documented here in detail). In the 5-dimensional case all vine structures are very similar, i.e.~a change of one edge can change an R-vine into a C- or D-vine.
	\item The Gauss model is the first detected model, i.e.~its size-power curve is the steepest and outperforms the other two. The C- and D-vine are more flexible in their choice of pair-copula families and thus can fit the data better.
	\item If the number of observations is too low, e.g.~$n=300$, conclusions are less robust. This weakness can be often observed in the inferential context and is e.g.~documented for other goodness-of-fit tests for copulas in the comparison study of \cite{Genest2009}.
	\item Very low dependencies yield to flatter size-power curves. If Kendall's $\tau$ is very small (in absolute terms) all copulas are close to the product copula and the choice of the vine structure as well as of the pair-copula families are less significant. 
	Increasing power by increasing strength of dependence are already observed in other goodness-of-fit tests for copulas, see e.g.~\cite{Genest2009} or \cite{Dufour2012}. 
	\item Additional simulation studies with 8-dimensional vine copula models show that the power decreases (slightly) for increasing dimension. With increasing dimension the number of pair-copulas and thus the number of copula parameters increases, e.g. in scenario 3 (Kendall's $\tau=0.5$) the power of an R-vine copula against a C-vine copula decreases at size 5\% from 92\% to 57\%, and against a Gauss copula from 78\% to 20\%, but against a D-vine copula it increases from 82\% to 89\%. For bigger size the gap shrinks. Our assumption is that numerical instabilities in the calculation of the gradient and especially of the Hessian matrix increase. Nevertheless, in 8 dimensions the power is still substantive.
\end{itemize}

\begin{table}[!ht]
\centering
\begin{tabular}{llrrrrrr}
  \toprule
True model &  & & & & & & \\
in the & Vine  & \multicolumn{2}{c}{$\tau=0.25$} & \multicolumn{2}{c}{$\tau=0.5$} & \multicolumn{2}{c}{mixed $\tau$} \\
alternative $H_1$ & under $H_0$ & Simul. & Asy. & Simul. & Asy. & Simul. & Asy.\\
  \midrule
R-vine  & R-vine & {\it 5.0} & {\it 31.9} &  {\it 5.0} & {\it 53.5} &  {\it 5.0} & {\it 44.4} \\
				& C-vine & 18.4 & 59.9 & 92.8 & 100.0 & 42.5 & 89.8 \\
  			& D-vine & 15.6 & 56.8 & 82.7 & 98.2 & 40.8 & 90.1 \\ 
  			& Gauss	 & 38.7 & 81.9 & 78.1 & 99.6 & 33.2 & 88.5 \\ 
\addlinespace 
C-vine	& R-vine & 15.8 & 54.6 & 59.8 & 97.7 & 30.8 & 88.0 \\
				& C-vine & {\it 5.0} & {\it 29.3} & {\it 5.0} & {\it 47.8} & {\it 5.0} & {\it 46.3} \\    
  			& D-vine & 14.1 & 50.5 & 67.6 & 98.2 & 51.8 & 94.1 \\  
  			& Gauss  & 17.3 & 57.6 & 58.4 & 97.3 & 36.5 & 92.1 \\
\addlinespace 
D-vine	& R-vine &  6.8 & 36.1 & 35.8 & 93.8 & 54.3 & 95.5 \\   
   			& C-vine &  9.4 & 41.6 & 62.3 & 98.8 & 35.8 & 90.0 \\
   			& D-vine & {\it 5.0} & {\it 29.0} & {\it 5.0} & {\it 51.2} & {\it 5.0} & {\it 44.4} \\   
  			& Gauss  & 17.5 & 56.5 & 37.9 & 94.3 & 60.4 & 96.5 \\ 
\addlinespace
Gauss   & R-Vine &  6.1 & 9.1 & 7.3 & 28.3 & 7.7 & 17.9 \\
  		 	& C-vine &  6.1 & 9.3 & 6.3 & 25.4 & 6.6 & 16.0 \\ 
   			& D-vine &  5.8 & 8.4 & 8.0 & 30.3 & 7.4 & 17.5 \\
   			& Gauss	 & {\it 5.0} & {\it 7.6} & {\it 5.0} & {\it 23.1} & {\it 5.0} & {\it 13.2} \\
\bottomrule
\end{tabular}
\caption{Estimated power (in \%) for $n=500$ at \textit{nominal} size 5\% (Values in italic give the actual size of the test)}
\label{tab:powerTable}
\end{table}

\subsubsection{Power study II}
\begin{figure}[hbtp]
	\centering
		\includegraphics[width=0.80\textwidth]{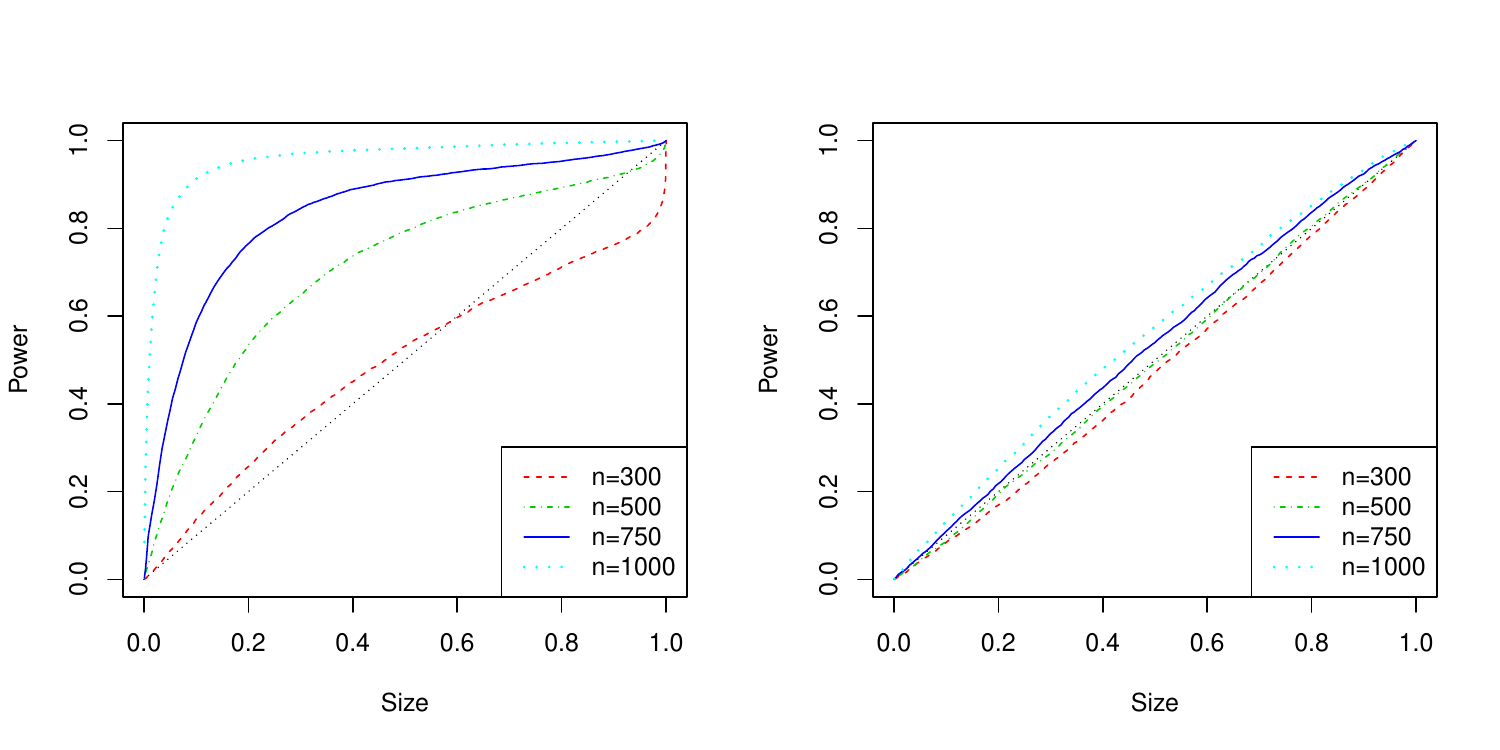} 
	\caption{Simulated size-power curves for different sample sizes in power study II. Left panel: R-vine model fitted with maximum spanning tree (MST) and sequential estimation, right panel: R-vine model fitted with Bayesian approach.}
	\label{fig:powerstudy_Lutz}
\end{figure}

\noindent
In the second scenario we investigate if the test can distinguish between two different 5-dimensional R-vines. One is selected and estimated with a maximum spanning tree (MST) algorithm and sequential estimation based on bivariate maximum likelihood (MLE), which is quite close to the full MLE \citep{DissmannBrechmannCzadoKurowicka2011}. The comparison candidate is an R-vine chosen via MCMC in a Bayesian approach with the highest probability among the visited variants in the MCMC \citep{GruberCzado2012}.
The true data generating model and the specification of the two estimated models are given in Table \ref{tab:5dimRvineScenario2} of  \ref{appendix:powerstudy2}. The model selected by the Bayesian approach differs from the original R-vine just in two copulas, i.e.~$c_{2,4}$ instead of $c_{3,4}$ and consequently $c_{3,4|2}$ instead of $c_{2,4|3}$. In contrast, the MST model differs in its fitted structure $\hat{\mathcal{V}}$ and pair-copula family selection $\hat{\mathcal{B}}(\hat{\mathcal{V}})$ to the other two more pronounced. 
Figure \ref{fig:powerstudy_Lutz} shows the results and the conclusions are:
\begin{itemize}
	\item The MST model is clearly detected as different from the true R-vine model for $n>500$, while the MCMC model, which is much "closer" to the true model, since the size-power curves are close to 45 degree for all sample sizes $n$.
	\item As in power study I the power for the MST model is increasing with increasing number of observations.
	\item The observed log-likelihood is a first indicator for the misspecification in the MST case, since $l_{MST} = 3360 < l_{MCMC} = 3731 < l_{true} = 3757$ (n=1000)\footnote{Results from \cite{GruberCzado2012}}. The log-likelihood of the MCMC model is much closer to the true one.
\end{itemize}

\subsubsection{Power study III}
\begin{figure}[htb]
	\centering
		\includegraphics[width=0.90\textwidth]{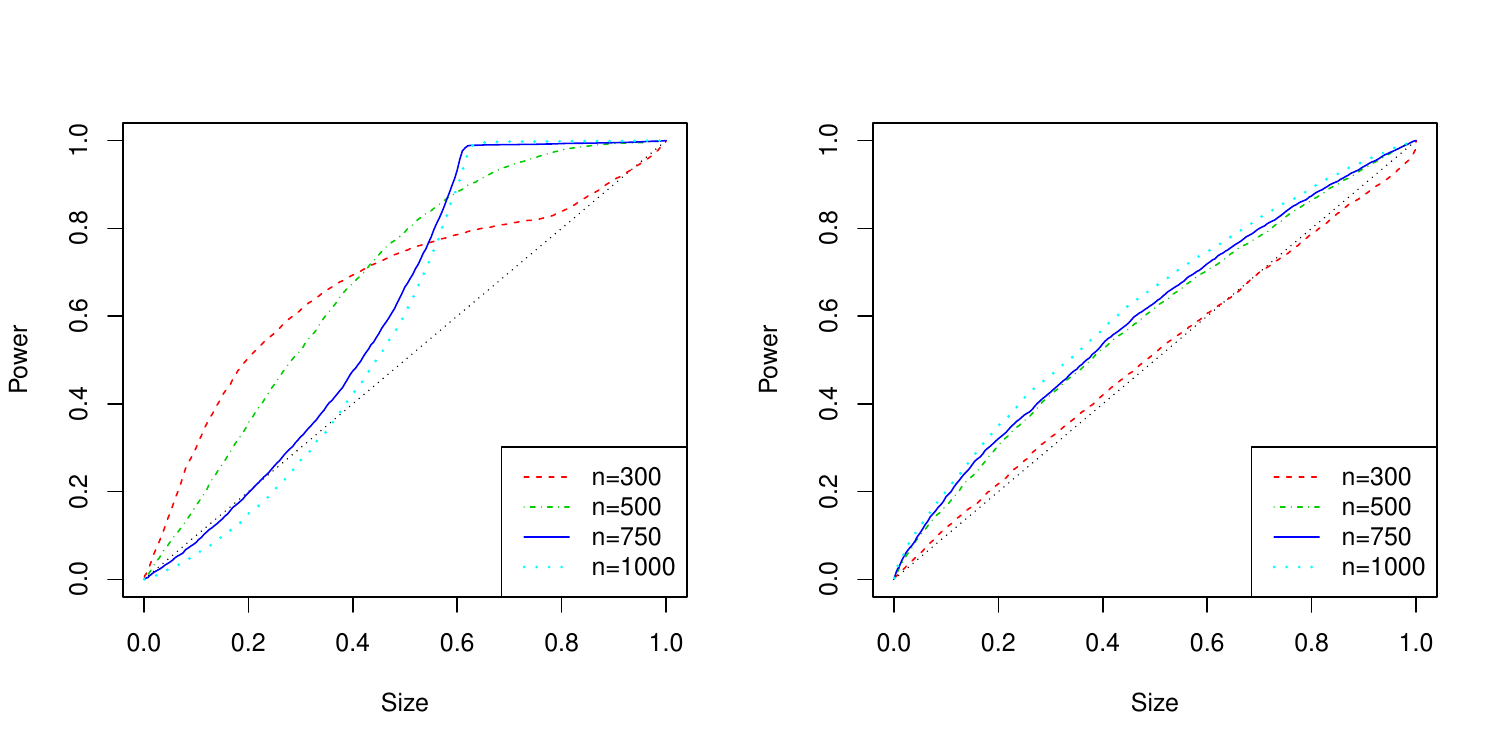}
	\caption{Size-power curves based on simulated p-values for different sample sizes. Left: R-vine (vine in $H_1$) versus multivariate t-copula (vine under $H_0$); right: multivariate t-copula (vine in $H_1$) versus R-vine (vine under $H_0$).}
	\label{fig:powerstudy_tCopula}
\end{figure}

\noindent
In financial applications often the multivariate Student's t-distribution or multivariate t-copula is used. In the last simulation study we investigate the difference of the multivariate t-copula with common degree-of-freedom versus an R-vine copula based on bivariate t-copulas with variable and separately estimated degrees-of-freedom in terms of our gof test. Again we choose the mixed R-vine model from power study I but change all copula families to t-copulas with the degree-of-freedom parameters in the range $\nu\in[4,20]$. \\
Using the convergence of the t-copula for large $\nu$ to the Gauss copula we replace t-copulas with estimated degrees-of-freedom greater or equal 30 with the Gaussian copula in the R-vine copula model. This is done for stability and accuracy reasons in the calculations of the derivatives.\\
In Figure \ref{fig:powerstudy_tCopula} we give the size-power curves for the simulated p-values for the two scenarios. 
The right panel indicates that given an R-vine with t-copulas the multivariate t-copula is not a good fit, since the test has power to discriminate against the multivariate t-copula. In contrast, the test has less power to discriminate an R-vine as alternative to a multivariate t-copula regardless of the sample size. For $n=300$ the test has no detection power since the size-power curves are close to the 45 degree line.\\
Further observations are:
\begin{itemize}
	\item The shape of the size-power curves depends heavily on the sample size \\
The numerical instability in estimation of the degrees-of-freedom may be a reason for this behavior.
	\item Compared to the other scenarios we have less power for small size.
\end{itemize}

\section{Extension to unknown margins}
\label{sec:extension}

\noindent
Goodness-of-fit tests such as the tests discussed \cite{Berg} do not account for uncertainties in the margins. He uses rank transformed data, i.e.~so called pseudo-samples $\boldsymbol{u}_1=(u_{11},\ldots,u_{1d}),\ldots,\boldsymbol{u}_n=(u_{n1},\ldots,u_{nd})$, where 
\[
	\boldsymbol{u}_j = (u_{j1},\ldots,u_{jd}) := \left( \frac{R_{j1}}{n+1},\ldots,\frac{R_{jd}}{n+1} \right),
\]
and $R_{ji}$ is the rank of $x_{ij}, i=1,\ldots,n$ and $j=1,\ldots,d$, which are i.i.d.~observations of the random vector $\bX=(X_1,\ldots,X_d)$. The denominator $n+1$ instead of $n$ avoids numerical problems at the boundaries of $[0,1]$.\\
A second method to handle unknown margins is the inference functions for margins (IFM) approach by \cite{Joe}, which uses parametric estimates $F_{\hat{\gamma}_i}, i=1,\ldots,d$ of the margins based on marginal parameter estimates $\hat{\gamma}_i$ and use them to transform to copula data. \\
In the case of unknown margins one has to adjust the computation of the test statistic. \cite{HuangProkhorov2011} did this for their copula gof test. Similarly we can adjust our proposed R-vine copula goodness-of-fit test.\\
The asymptotic variance matrix $V_{\btheta_0}$ (Expression (\ref{eq:varianceMatrix})) for the test statistic $\calT_n$ (Expression (\ref{eq:teststatistic})) is extended using expected derivatives with respect to the margins of the log-likelihood and the expected derivatives of the vectorized sum of the Hessian matrix and the outer product of gradient, respectively. More precisely define
{\small
\begin{equation*}
\begin{split}
	W_i(F_{i}) &:= \int_{[0,1]^d}\left[ I_{\{F_{i}\leq u_i\}}-u_i \right] \partial_{\btheta,u_i}^2 \ln(c_{\btheta_0}(u_1,\ldots,u_d)) c_{\btheta_0}(u_1,\ldots,u_d)du_1\ldots du_d, \\
	M_i(F_{i}) &:= \int_{[0,1]^d}\left[ I_{\{F_{i}\leq u_i\}}-u_i \right] \partial_{u_i} vech\bigg(\partial_{\btheta}^2 \ln(c_{\btheta_0}(u_1,\ldots,u_d)) +  \\ 
	&\qquad \partial_{\btheta} \ln(c_{\btheta_0}(u_1,\ldots,u_d))\big(\partial_{\btheta} \ln(c_{\btheta_0}(u_1,\ldots,u_d))\big)^T \bigg) du_1\ldots du_d,
\end{split}
\end{equation*}}\noindent
with $F_{i}:=F_i(x_{i}), i=1,\ldots,d$. Furthermore, $\boldsymbol{d}(\btheta_0)$ is now defined in terms of the random vector $\bX$:
\[
	\boldsymbol{d}(\btheta_0|\bX):=vech(\bbh(\btheta_0)+\bbc(\btheta_0)),
\]
where $\bbh(\btheta_0)$ and $\bbc(\btheta_0)$ are defined in (\ref{eq:HC}). With $l(\btheta|\bX):=\ln(c_{\btheta_0}(F_1,\ldots,F_d))$ the adjusted variance matrix is
{\footnotesize
\begin{equation*}
\begin{split}
	V_{\btheta_0} = E\Bigg[ &\left(\boldsymbol{d}(\btheta_0|\bX) - \nabla D_{\btheta_0}\bbh^{-1}(\btheta_0)\left( \partial_{\btheta}l(\btheta_0|\bX) + \sum_{i=1}^d W_i(F_{i}) \right) + \sum_{i=1}^d M_i(F_{i}) \right)  \\
	&\left. \left(\boldsymbol{d}(\btheta_0|\bX) - \nabla D_{\btheta_0}\bbh^{-1}(\btheta_0)\left( \partial_{\btheta}l(\btheta_0|\bX) + \sum_{i=1}^d W_i(F_{i}) \right) + \sum_{i=1}^d M_i(F_{i}) \right)^T
	\right].
\end{split}
\label{eq:V0}
\end{equation*}}

\begin{figure}[hbtp]
	\centering
		\includegraphics[width=0.95\textwidth]{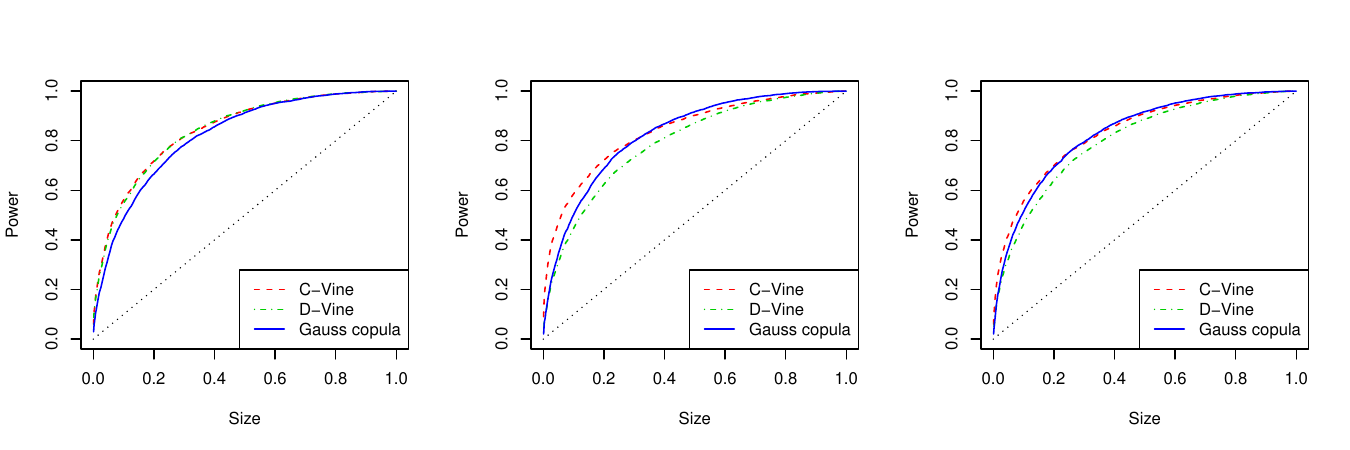} 
	\caption{Simulated size-power curves for the mixed Kendall's $\tau$ R-vine copula model considering $n=500$ observations. left panel: no uncertainty in the margins; middle panel: data generating process with standard normal margins for all 5 dimensions, right panel: data generating process with centered normal margins with different standard deviations.}
	\label{fig:powerstudyMargins}
\end{figure}

\noindent
But, this correction of White's original formula would involve multidimensional integrals, not computational tractable in appropriate time. An adjusted gradient and Hesse matrix may avoid this problem. Since the goodness-of-fit test calculation does not depend directly on the density function $f$ but on the product of pair-copulas and marginal density functions, the derivatives should not only be with respect to the parameters but to the marginals too. This approach may be a topic for further research. Thus (\ref{eq:varianceMatrix}) will be used as an approximation in the case of unknown margins, e.g.~in our application in the next section.\\
To justify the good approximation behavior of (\ref{eq:varianceMatrix}) we run power study I of Section \ref{sec:powerstudy} with unknown margins. We limit ourselves here to the mixed copula case with $n=500$ observations. The chosen DGP uses the standard normal distribution for all 5 margins in a first scenario, and centered normal distributions with different standard deviations $\sigma\in\{1,2,3,4,5\}$ in a second scenario. The margins are estimated via moments in an IFM approach.\\
In Figure \ref{fig:powerstudyMargins} we illustrate the resulting size-power curves. Comparing the middle panel to and the right panel to the left panel we cannot detect significant differences in the power if the margins are unknown. Further, the choice of the marginal distribution $F_{\gamma}$ has no significant influence on the size-power curves. This is confirmed by further power studies not presented in this manuscript, e.g.~marginal Student's t-distributions. This is no longer true if the choice of the marginal distribution fits badly the (generated) data, e.g.~data generated from a Student's t-distribution fitted with an exponential distribution. An estimation of the margins with the rank based approach returned similar results as the IFM approach above.

\section{Application}
\label{sec:application}

\noindent
Finally, we apply our introduced goodness-of-fit test to a financial data set in order to investigate different fitted vine models. The considered data are 8 daily US-exchange rates covering the time horizon from July 22, 2005 to July 17, 2009, resulting in 1007 data points in total. First of all it was discussed in \cite[Chapter 5]{Schepsmeier} and \cite{CzadoSchepsmeierMin2011} considering a C-vine and later in \cite{stoeber2011c} in a regime switching model with R-vines and in a rolling window analysis with an R-vine in \cite{StoeberSchepsmeier2012}.\\
Each marginal exchange rate time series is fitted with an appropriate AR(1)-GARCH(1,1) model with skewed t innovations. The resulting standardized residuals are transformed using the non-parametric rank transformation (see \citealt{Genest1995}) to obtain $[0,1]^8$ copula data. For details on the modeling of the margins we refer to \cite{StoeberSchepsmeier2012}. For the notation of the variables we follow \cite{StoeberSchepsmeier2012} with 1=AUD (Australian dollar), 2=JPY
(Japanese yen), 3=BRL (Brazilian real), 4=CAD (Canadian dollar), 5=EUR (Euro), 6=CHF (Swiss frank), 7=INR (Indian rupee) and 8=GBP (British pound). In \cite{Schepsmeier} and \cite{CzadoSchepsmeierMin2011} a slightly different notation were used.\\

\noindent
Performing a parametric bootstrap with repetition rate $B=1000$ and sample size $N=5000$ our goodness-of-fit test results (see Table \ref{tab:indices}) confirm that the C-vine model of \cite{CzadoSchepsmeierMin2011} can not be rejected at a 5\% significance level, i.e. that the C-vine fits the data quite well. The R-vine model of \cite{StoeberSchepsmeier2012} has only a bootstrapped p-value of 1\%, i.e. has a smaller significance than the C-vine but is still a model to be favoured over the frequently used Gauss copula, which as a p-value of 0.
The log-likelihood, AIC and BIC show a similar picture since the C-vine is preferred in terms of log-likelihood and AIC. Looking at the BIC criterium the R-vine is favorable due to its relative small number of parameters and a quite high log-likelihood.
Also, our goodness-of-fit test is consistent with the previous findings that the more flexible R-vine and C-vine are appropriate for financial modeling.

\begin{table}[ht]
\begin{center}
\begin{tabular}{lrrrrrr}
  \toprule
model & loglik & AIC & BIC & $\#par$ & $\calT$ & p-value \\ 
  \midrule
C-vine 				&  \textbf{2213} & \textbf{-4358} & -4191 & 34  & 705 & \textbf{0.63} \\ 
R-vine 				&  2199 & -4343 & \textbf{-4205} & 28 & 493 & 0.01 \\ 
mult. Gauss 	&  2089 & -4121 & -3984 & 28 & 365 & 0 \\ 
   \bottomrule
\end{tabular}
\caption{Log-likelihood, AIC, BIC, number of parameters, and the test statistics and bootstrapped p-values of the goodness-of-fit test for the investigated models.}
\label{tab:indices}
\end{center}
\end{table}

\section{Summary and discussion}
\label{sec:discussion}

\noindent
We introduced the first goodness-of-fit test for R-vine copula models based on the information matrix equality and specification test proposed by \cite{White1982}. It extends the approach of \cite{HuangProkhorov2011} to the far more flexible class of regular vine copulas. Thus we have a further comparison tool for vine copula model selection. Goodness-of-fit tests for copulas are broadly discussed and compared in the literature, but mainly for bivariate copulas. The vine copula constructions lacked so far for a suitable test. \\
Especially, the possibility to compare the flexible class of vine copulas to the often used multivariate Gauss and Student's t copula is welcome.\\
The proposed test's advantages are its simplicity, easy implementation (having the derivatives) and asymptotic distribution.
The extensive power study showed the test's good power behavior against wrong specified models given simulated p-values.\\
But this is also one of its greatest criticism points: The test has a poor behavior in small sample sizes given asymptotic p-values. The asymptotic based test is inaccurate for sample sizes smaller than 10000 in 5 dimensions. In higher dimensions even higher sample sizes are needed. Furthermore, in an undercharged sample the test does not reach its \textit{nominal} size given asymptotic p-values.\\ 
Also the increasing power with increasing sample size can be detected in many other tests and is often observed in the inferential context.\\
If the margins are unknown we had to simplify the calculation of the variance matrix $V_0$ in the test statistic evaluation by an approximated version. Thus strong statements and interpretations in this case should be avoided. In future work this weakness will be closer investigated. An idea might be the extension of the Hessian matrix and the outer-product of the score function by the derivatives with respect to the margins. A comparison study with other suggested but not investigated goodness-of-fit tests for vine copulas such as tests based on PIT (suggested in \cite{Aas_Czado}), the empirical copula or Kendall's transform (used in \cite{BergAas2009} for a 4-dimensional C-vine) will be done in the future.

\section{Acknowledgment}
\noindent
The author acknowledge substantial contributions by his colleagues of the research group of Prof. C. Czado at Technische Universit\"at M\"unchen and the the support of the TUM Graduate School's International School of Applied Mathematics. Further thanks goes to Wanling Huang who supported me with code provided by personal communication. Numerical calculations were performed on a Linux cluster supported by DFG grant INST 95/919-1 FUGG.

\bibliography{references}

\begin{thebibliography}{37}
\expandafter\ifx\csname natexlab\endcsname\relax\def\natexlab#1{#1}\fi
\expandafter\ifx\csname url\endcsname\relax
  \def\url#1{\texttt{#1}}\fi
\expandafter\ifx\csname urlprefix\endcsname\relax\def\urlprefix{URL }\fi
\providecommand{\eprint}[2][]{\url{#2}}
\providecommand{\bibinfo}[2]{#2}
\ifx\xfnm\relax \def\xfnm[#1]{\unskip,\space#1}\fi
\bibitem[{Aas et~al.(2009)Aas, Czado, Frigessi and Bakken}]{Aas_Czado}
\bibinfo{author}{Aas, K.}, \bibinfo{author}{Czado, C.},
  \bibinfo{author}{Frigessi, A.}, \bibinfo{author}{Bakken, H.},
  \bibinfo{year}{2009}.
\newblock \bibinfo{title}{Pair-copula construction of multiple dependence}.
\newblock \bibinfo{journal}{Insurance: Mathematics and Economics}
  \bibinfo{volume}{44}, \bibinfo{pages}{182--198}.
\bibitem[{Bedford and Cooke(2001)}]{Bedford_Cooke2001}
\bibinfo{author}{Bedford, T.}, \bibinfo{author}{Cooke, R.},
  \bibinfo{year}{2001}.
\newblock \bibinfo{title}{Probability density decomposition for conditionally
  dependent random variables modeled by vines.}
\newblock \bibinfo{journal}{Annals of Mathematics and Artificial Intelligence}
  \bibinfo{volume}{32}, \bibinfo{pages}{245--268}.
\bibitem[{Bedford and Cooke(2002)}]{Bedford_Cooke}
\bibinfo{author}{Bedford, T.}, \bibinfo{author}{Cooke, R.},
  \bibinfo{year}{2002}.
\newblock \bibinfo{title}{Vines - a new graphical model for dependent random
  variables.}
\newblock \bibinfo{journal}{Annals of Statistics} \bibinfo{volume}{30},
  \bibinfo{pages}{1031--1068}.
\bibitem[{Berg(2009)}]{Berg}
\bibinfo{author}{Berg, D.}, \bibinfo{year}{2009}.
\newblock \bibinfo{title}{Copula goodness-of-fit testing: An overview and power
  comparison}.
\newblock \bibinfo{journal}{The European Journal of Finance}
  \bibinfo{volume}{15}, \bibinfo{pages}{1466--4364}.
\bibitem[{Berg and Aas(2009)}]{BergAas2009}
\bibinfo{author}{Berg, D.}, \bibinfo{author}{Aas, K.}, \bibinfo{year}{2009}.
\newblock \bibinfo{title}{Models for construction of multivariate dependence: A
  comparison study}.
\newblock \bibinfo{journal}{The European Journal of Finance}
  \bibinfo{volume}{15}, \bibinfo{pages}{639--659}.
\bibitem[{Brechmann and Czado(2013)}]{BrechmannEuroStoxx2012}
\bibinfo{author}{Brechmann, E.}, \bibinfo{author}{Czado, C.},
  \bibinfo{year}{2013}.
\newblock \bibinfo{title}{{Risk Management with High-Dimensional Vine Copulas:
  An Analysis of the Euro Stoxx 50.}}
\newblock \bibinfo{journal}{Statistics \& Risk Modeling} \bibinfo{note}{To
  appear}.
\bibitem[{Brechmann et~al.(2012)Brechmann, Czado and
  Aas}]{BrechmannCzadoAas2012}
\bibinfo{author}{Brechmann, E.}, \bibinfo{author}{Czado, C.},
  \bibinfo{author}{Aas, K.}, \bibinfo{year}{2012}.
\newblock \bibinfo{title}{Truncated regular vines in high dimensions with
  applications to financial data}.
\newblock \bibinfo{journal}{Canadian Journal of Statistics}
  \bibinfo{volume}{40}, \bibinfo{pages}{68--85}.
\bibitem[{Brechmann and Schepsmeier(2013)}]{BrechmannSchepsmeier2011}
\bibinfo{author}{Brechmann, E.C.}, \bibinfo{author}{Schepsmeier, U.},
  \bibinfo{year}{2013}.
\newblock \bibinfo{title}{{Dependence modeling with C- and D-vine copulas: The
  R-package CDVine}}.
\newblock \bibinfo{journal}{Journal of Statistical Software}
  \bibinfo{volume}{52}, \bibinfo{pages}{1--27}.
\bibitem[{Breymann et~al.(2003)Breymann, Dias and Embrechts}]{Breymann2003}
\bibinfo{author}{Breymann, W.}, \bibinfo{author}{Dias, A.},
  \bibinfo{author}{Embrechts, P.}, \bibinfo{year}{2003}.
\newblock \bibinfo{title}{Dependence structures for multivariate high-frequency
  data in finance}.
\newblock \bibinfo{journal}{Quantitative Finance} \bibinfo{volume}{3},
  \bibinfo{pages}{1--14}.
\bibitem[{Chesher and Spady(1991)}]{ChesherSpady1991}
\bibinfo{author}{Chesher, A.}, \bibinfo{author}{Spady, R.},
  \bibinfo{year}{1991}.
\newblock \bibinfo{title}{Asymptotic expansions of the information matrix test
  statistic}.
\newblock \bibinfo{journal}{Econometrica} \bibinfo{volume}{59},
  \bibinfo{pages}{787--815}.
\bibitem[{Clarke(2007)}]{Clarke}
\bibinfo{author}{Clarke, K.}, \bibinfo{year}{2007}.
\newblock \bibinfo{title}{{A Simple Distribution-Free Test for Nonnested Model
  Selection}}.
\newblock \bibinfo{journal}{Political Analysis} \bibinfo{volume}{15},
  \bibinfo{pages}{347--363}.
\bibitem[{Czado(2010)}]{Czado}
\bibinfo{author}{Czado, C.}, \bibinfo{year}{2010}.
\newblock \bibinfo{title}{{Pair-Copula Constructions of Multivariate Copulas}},
  in: \bibinfo{editor}{{Jaworski, P. and Durante, F. and H\"ardle, W.K. and
  Rychlik, T}} (Ed.), \bibinfo{booktitle}{{ Copula Theory and Its Applications,
  Lecture Notes in Statistics}}, \bibinfo{publisher}{Springer-Verlag},
  \bibinfo{address}{Berlin Heidelberg}. pp. \bibinfo{pages}{93--109}.
\bibitem[{Czado et~al.(2012)Czado, Schepsmeier and
  Min}]{CzadoSchepsmeierMin2011}
\bibinfo{author}{Czado, C.}, \bibinfo{author}{Schepsmeier, U.},
  \bibinfo{author}{Min, A.}, \bibinfo{year}{2012}.
\newblock \bibinfo{title}{Maximum likelihood estimation of mixed {C}-vines with
  application to exchange rates}.
\newblock \bibinfo{journal}{Statistical Modelling} \bibinfo{volume}{12},
  \bibinfo{pages}{229--255}.
\bibitem[{Davidson and MacKinnon(1998)}]{DavidsonMacKinnon1998}
\bibinfo{author}{Davidson, R.}, \bibinfo{author}{MacKinnon, J.},
  \bibinfo{year}{1998}.
\newblock \bibinfo{title}{Graphical methods for investigating the size and
  power of hypothesis tests}.
\newblock \bibinfo{journal}{The Manchester School} \bibinfo{volume}{66},
  \bibinfo{pages}{1--26}.
\bibitem[{Dißmann et~al.(2013)Dißmann, Brechmann, Czado and
  Kurowicka}]{DissmannBrechmannCzadoKurowicka2011}
\bibinfo{author}{Dißmann, J.}, \bibinfo{author}{Brechmann, E.},
  \bibinfo{author}{Czado, C.}, \bibinfo{author}{Kurowicka, D.},
  \bibinfo{year}{2013}.
\newblock \bibinfo{title}{Selecting and estimating regular vine copulae and
  application to financial returns}.
\newblock \bibinfo{journal}{Computational Statistics and Data Analysis}
  \bibinfo{volume}{59}, \bibinfo{pages}{52 -- 69}.
\bibitem[{Dufour et~al.(2012)Dufour, Genest and Huang}]{Dufour2012}
\bibinfo{author}{Dufour, J.M.}, \bibinfo{author}{Genest, C.},
  \bibinfo{author}{Huang, W.}, \bibinfo{year}{2012}.
\newblock \bibinfo{title}{A regularized goodness-of-fit test for copulas}.
\newblock \bibinfo{note}{In revision, Personal communication}.
\bibitem[{Fawcett(2006)}]{Fawcett2006}
\bibinfo{author}{Fawcett, T.}, \bibinfo{year}{2006}.
\newblock \bibinfo{title}{{An introduction to ROC analysis}}.
\newblock \bibinfo{journal}{Pattern Recognition Letters} \bibinfo{volume}{27},
  \bibinfo{pages}{861--874}.
\bibitem[{Genest et~al.(1995)Genest, Ghoudi and Rivest}]{Genest1995}
\bibinfo{author}{Genest, C.}, \bibinfo{author}{Ghoudi, K.},
  \bibinfo{author}{Rivest, L.P.}, \bibinfo{year}{1995}.
\newblock \bibinfo{title}{A semiparametric estimation procedure of dependence
  parameters in multivariate families of distributions}.
\newblock \bibinfo{journal}{Biometrika} \bibinfo{volume}{82},
  \bibinfo{pages}{543--552}.
\bibitem[{Genest et~al.(2006)Genest, Quessy and
  Re{\'e}millard}]{Genest_Remillard_2}
\bibinfo{author}{Genest, C.}, \bibinfo{author}{Quessy, J.F.},
  \bibinfo{author}{Re{\'e}millard, B.}, \bibinfo{year}{2006}.
\newblock \bibinfo{title}{{Goodness-of-fit Procedures for Copula Model Based on
  the Probability Integral Transformation}}.
\newblock \bibinfo{journal}{{Scandinavian Journal of Statistics}}
  \bibinfo{volume}{33}, \bibinfo{pages}{337--366}.
\bibitem[{Genest et~al.(2009)Genest, R{\'e}millard and Beaudoin}]{Genest2009}
\bibinfo{author}{Genest, C.}, \bibinfo{author}{R{\'e}millard, B.},
  \bibinfo{author}{Beaudoin, D.}, \bibinfo{year}{2009}.
\newblock \bibinfo{title}{Goodness-of-fit tests for copulas: a review and power
  study}.
\newblock \bibinfo{journal}{Insurance: Mathematics and Economics}
  \bibinfo{volume}{44}, \bibinfo{pages}{199--213}.
\bibitem[{Gruber and Czado(2013)}]{GruberCzado2012}
\bibinfo{author}{Gruber, L.}, \bibinfo{author}{Czado, C.},
  \bibinfo{year}{2013}.
\newblock \bibinfo{title}{{Sequential Bayesian Model Selection of Regular Vine
  Copulas}}.
\newblock \bibinfo{note}{Submitted for publication, Personal communication}.
\bibitem[{Hall(1989)}]{Hall1989}
\bibinfo{author}{Hall, A.}, \bibinfo{year}{1989}.
\newblock \bibinfo{title}{On the calculation of the information matrix test in
  the normal linear regression model}.
\newblock \bibinfo{journal}{Economics Letters} \bibinfo{volume}{29},
  \bibinfo{pages}{31--35}.
\bibitem[{Hob{\ae}k~Haff(2013)}]{haff2010}
\bibinfo{author}{Hob{\ae}k~Haff, I.}, \bibinfo{year}{2013}.
\newblock \bibinfo{title}{Parameter estimation for pair-copula constructions}.
\newblock \bibinfo{journal}{Bernoulli} \bibinfo{volume}{19},
  \bibinfo{pages}{462--491}.
\bibitem[{Huang and Prokhorov(2013)}]{HuangProkhorov2011}
\bibinfo{author}{Huang, W.}, \bibinfo{author}{Prokhorov, A.},
  \bibinfo{year}{2013}.
\newblock \bibinfo{title}{A goodness-of-fit test for copulas}.
\newblock \bibinfo{journal}{Economic Reviews} \bibinfo{note}{To appear}.
\bibitem[{Joe(1996)}]{Joe2}
\bibinfo{author}{Joe, H.}, \bibinfo{year}{1996}.
\newblock \bibinfo{title}{Families of m-variate distributions with given
  margins and m(m-1)/2 bivariate dependence parameters}, in:
  \bibinfo{editor}{{L. R{\"{u}}schendorf and B. Schweizer and M. D. Taylor}}
  (Ed.), \bibinfo{booktitle}{Distributions with Fixed Marginals and Related
  Topics}, \bibinfo{publisher}{Inst. Math. Statist.},
  \bibinfo{address}{Hayward, CA}. pp. \bibinfo{pages}{120--141}.
\bibitem[{Joe(1997)}]{Joe}
\bibinfo{author}{Joe, H.}, \bibinfo{year}{1997}.
\newblock \bibinfo{title}{Multivariate Models and Dependence Concepts}.
\newblock \bibinfo{publisher}{Chapman und Hall, London}.
\bibitem[{Min and Czado(2010)}]{MinCzado2010}
\bibinfo{author}{Min, A.}, \bibinfo{author}{Czado, C.}, \bibinfo{year}{2010}.
\newblock \bibinfo{title}{Bayesian inference for multivariate copulas using
  pair-copula constructions}.
\newblock \bibinfo{journal}{Journal of Financial Econometrics}
  \bibinfo{volume}{8}, \bibinfo{pages}{511--546}.
\bibitem[{Min and Czado(2012)}]{min:czado:2010:scomdy}
\bibinfo{author}{Min, A.}, \bibinfo{author}{Czado, C.}, \bibinfo{year}{2012}.
\newblock \bibinfo{title}{{SCOMDY} models based on pair-copula constructions
  with application to exchange rates}.
\newblock \bibinfo{note}{To appear in: Computational Statistics and Data
  Analysis}.
\bibitem[{Morales-N{\'a}poles(2010)}]{MoralesNapoles2010}
\bibinfo{author}{Morales-N{\'a}poles, O.}, \bibinfo{year}{2010}.
\newblock \bibinfo{title}{{Counting Vines}}, in: \bibinfo{editor}{Kurowicka,
  D.~Joe, H.} (Ed.), \bibinfo{booktitle}{{Dependence Modeling-Handbook on Vine
  Copulas}}, \bibinfo{publisher}{World Scientific Publishing},
  \bibinfo{address}{Singapore}. pp. \bibinfo{pages}{189--218}.
\bibitem[{Schepsmeier(2010)}]{Schepsmeier}
\bibinfo{author}{Schepsmeier, U.}, \bibinfo{year}{2010}.
\newblock \bibinfo{title}{{Maximum likelihood estimation of C-vine pair-copula
  constructions on bivariate copulas from different families}}.
\newblock \bibinfo{type}{Diploma thesis}. Center of Mathematical Sciences,
  Munich University of Technology. \bibinfo{address}{Garching bei München}.
\bibitem[{Schepsmeier et~al.(2012)Schepsmeier, Stoeber and
  Brechmann}]{VineCopula}
\bibinfo{author}{Schepsmeier, U.}, \bibinfo{author}{Stoeber, J.},
  \bibinfo{author}{Brechmann, E.C.}, \bibinfo{year}{2012}.
\newblock \bibinfo{title}{VineCopula: Statistical inference of vine copulas}.
\newblock \bibinfo{note}{R package version 1.0}.
\bibitem[{St\"ober and Czado(2013)}]{stoeber2011c}
\bibinfo{author}{St\"ober, J.}, \bibinfo{author}{Czado, C.},
  \bibinfo{year}{2013}.
\newblock \bibinfo{title}{Detecting regime switches in the dependence structure
  of high dimensional financial data}.
\newblock \bibinfo{journal}{Computational Statistics and Data Analysis}
  \bibinfo{note}{To appear}.
\bibitem[{St\"ober and Schepsmeier(2013)}]{StoeberSchepsmeier2012}
\bibinfo{author}{St\"ober, J.}, \bibinfo{author}{Schepsmeier, U.},
  \bibinfo{year}{2013}.
\newblock \bibinfo{title}{Estimating standard errors in regular vine copula
  models}.
\newblock \bibinfo{note}{Submitted for publication available at:
  http://arxiv.org/abs/1205.4841}.
\bibitem[{Taylor(1987)}]{Taylor1987}
\bibinfo{author}{Taylor, L.W.}, \bibinfo{year}{1987}.
\newblock \bibinfo{title}{The size bias of white's information matrix test}.
\newblock \bibinfo{journal}{Economics Letters} \bibinfo{volume}{24},
  \bibinfo{pages}{63--67}.
\bibitem[{Vuong(1989)}]{Vuong}
\bibinfo{author}{Vuong, Q.}, \bibinfo{year}{1989}.
\newblock \bibinfo{title}{{Likelihood Ratio Tests for Model Selection and
  Non-Nested Hypotheses}}.
\newblock \bibinfo{journal}{Econometrica} \bibinfo{volume}{57},
  \bibinfo{pages}{307--333}.
\bibitem[{White(1982)}]{White1982}
\bibinfo{author}{White, H.}, \bibinfo{year}{1982}.
\newblock \bibinfo{title}{Maximum likelihood estimation of misspecified
  models}.
\newblock \bibinfo{journal}{Econometrica} \bibinfo{volume}{50},
  \bibinfo{pages}{1--26}.
\bibitem[{Yan(2007)}]{copula}
\bibinfo{author}{Yan, J.}, \bibinfo{year}{2007}.
\newblock \bibinfo{title}{Enjoy the joy of copulas: With a package {copula}}.
\newblock \bibinfo{journal}{Journal of Statistical Software}
  \bibinfo{volume}{21}, \bibinfo{pages}{1--21}.

\end{thebibliography}

\appendix
\section{Model specification in power study I}
\label{appendix:powerstudy}

\begin{gather}
\begin{split}
	f_{R}(x_1,&x_2,x_3,x_4,x_5) = f_1(x_1)f_2(x_2)f_3(x_3)f_4(x_4)f_5(x_5) \\
	&\cdot c_{1,2}\cdot c_{1,3}\cdot c_{1,4}\cdot c_{4,5}\cdot c_{2,4|1}\cdot c_{3,4|1}\cdot c_{1,5|4}\cdot 
 c_{2,3|1,4}\cdot c_{3,5|1,4}\cdot c_{2,5|1,3,4}
\end{split}
 \label{eq:5dimRvine}
\end{gather}
\begin{gather}
\begin{split}
	f_{C}(x_1,&x_2,x_3,x_4,x_5) = f_1(x_1)f_2(x_2)f_3(x_3)f_4(x_4)f_5(x_5) \\
	&\cdot c_{1,2}\cdot c_{2,3}\cdot c_{2,4}\cdot c_{2,5}\cdot c_{1,3|2}\cdot c_{1,4|2}\cdot c_{1,5|2}\cdot 
 c_{3,4|1,2}\cdot c_{4,5|1,2}\cdot c_{3,5|1,2,4}
\end{split}
\label{eq:5dimCvine}
\end{gather}
\begin{gather}
\begin{split}
	f_{D}(x_1,&x_2,x_3,x_4,x_5) = f_1(x_1)f_2(x_2)f_3(x_3)f_4(x_4)f_5(x_5) \\
	&\cdot c_{1,2}\cdot c_{1,5}\cdot c_{4,5}\cdot c_{3,4}\cdot c_{2,5|1}\cdot c_{1,4|5}\cdot c_{3,5|4}\cdot 
 c_{2,4|1,5}\cdot c_{1,3|4,5}\cdot c_{2,3|1,4,5}
\end{split}
 \label{eq:5dimDvine}
\end{gather}

\begin{figure}[htbp]
	\centering
		\includegraphics[width=0.90\textwidth]{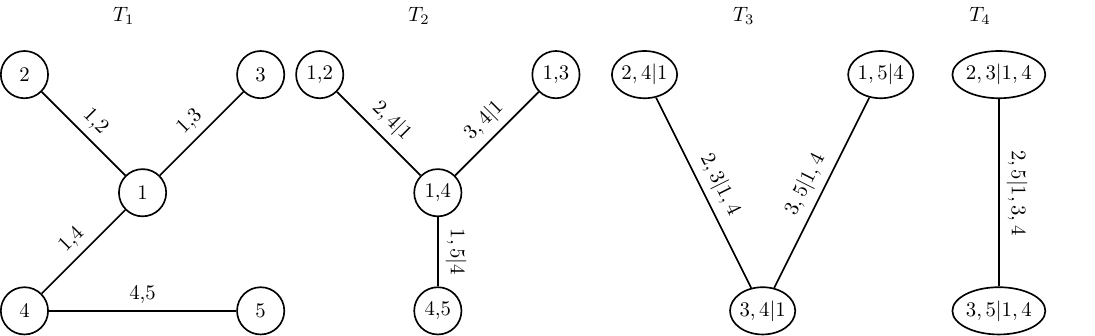}
	\caption{Tree structure of the 5 dimensional R-vine copula used in the power studies.}
	\label{fig:5dimRvine}
\end{figure}

\newpage

\begin{table}[!ht]
	\centering
		\begin{tabular}{llcr}
		\toprule
			Tree & $\mathcal{V}_R$ & $\mathcal{B}_R(\mathcal{V}_R)$ & $\tau$  \\
			\midrule
			1 & $c_{1,2}$ & Gauss & 0.71 \\
				& $c_{1,3}$ & Gauss & 0.33 \\ 
				& $c_{1,4}$ & Clayton & 0.71 \\
				& $c_{4,5}$ & Gumbel & 0.74 \\
			2	& $c_{2,4|1}$ & Gumbel & 0.38 \\
				& $c_{3,4|1}$ & Gumbel & 0.47 \\
				& $c_{1,5|4}$ & Gumbel & 0.33 \\
			3	& $c_{2,3|1,4}$ & Clayton & 0.35 \\
				& $c_{3,5|1,4}$ & Clayton & 0.31 \\
			4 & $c_{2,5|1,3,4}$ & Gauss & 0.13 \\
			\bottomrule
		\end{tabular}
		\caption{Copula families and Kendall's $\tau$ values of the investigated (mixed) R-vine copula model defined by (\ref{eq:5dimRvine}).}
		\label{tab:5dimRvine}
\end{table}

\section{Model specification in power study II}
\label{appendix:powerstudy2}

\begin{table}[!ht]
	\centering
		\begin{tabular}{lcr|lc|lc}
		\toprule
			\multicolumn{3}{c}{True model ($M_1$)} & \multicolumn{2}{c}{$M_2^{MST}$} & \multicolumn{2}{c}{$M_2^{MCMC}$}\\
			\midrule
			$\mathcal{V}$ & $\mathcal{B}(\mathcal{V})$ & $\tau$ & $\hat{\mathcal{V}}$ & $\hat{\mathcal{B}}(\hat{\mathcal{V}})$ & $\hat{\mathcal{V}}$ & $\hat{\mathcal{B}}(\hat{\mathcal{V}})$ \\
			\midrule
			$c_{1,2}$ & Gauss & 0.10 & $c_{1,3}$ & $t_{\nu}$ & $c_{1,2}$ & Gauss \\
			$c_{2,3}$ & $t_3$ & -0.15 & $c_{1,5}$ & Gauss & $c_{2,3}$ & $t_{\nu}$ \\
			$c_{3,4}$ & $t_3$ & -0.10 & $c_{2,5}$ & $t_{\nu}$ & $c_{2,4}$ & Gumbel 90 \\
			$c_{3,5}$ & $t_3$ & 0.15 & $c_{4,5}$ & Gumbel 270 & $c_{3,5}$ & $t_{\nu}$ \\
			$c_{1,3|2}$ & N & 0.70 & $c_{1,2|5}$ & $t_{\nu}$ & $c_{1,3|2}$ & Gauss \\
			$c_{2,4|3}$ & Gumbel 90 & -0.60 & $c_{1,4|5}$ & $t_{\nu}$ & $c_{3,4|2}$ & Gumbel \\
			$c_{2,5|3}$ & Gumbel & 0.85 & $c_{3,5|1}$ & $t_{\nu}$ & $c_{2,5|3}$ & $t_{\nu}$ \\
			$c_{1,4|2,3}$ & Gauss & 0.45 & $c_{2,3|1,5}$ & $t_{\nu}$ & $c_{1,4|2,3}$ & Gauss \\
			$c_{1,5|2,3}$ & Gauss & -0.50 & $c_{3,4|1,5}$ & $t_{\nu}$ & $c_{1,5|2,3}$ & Gauss\\
			$c_{4,5|1,2,3}$ & Gauss & 0.10 & $c_{2,4|1,3,5}$ & Gauss & $c_{4,5|1,2,3}$ & Gauss \\
			\bottomrule
		\end{tabular}
		\caption{Copula families and Kendall's $\tau$ values of the investigated R-vine models in power study 2 ($t_{\nu} \hat{=}$ t-copula with $\nu$ degrees-of-freedom, Gumbel 90 $\hat{=}$ 90 degree rotated Gumbel copula, Gumbel 270 $\hat{=}$ 270 degree rotated Gumbel copula).}
		\label{tab:5dimRvineScenario2}
\end{table}

\end{document}